\documentclass[twocolumn]{aastex6}

\fullcollaborationName{The Friends of AASTeX Collaboration}

\begin{document}

\title{A small-scale filament eruption inducing Moreton Wave, EUV Wave and Coronal Mass Ejection}

\author{Jincheng Wang\altaffilmark{1,2,3}, Xiaoli Yan\altaffilmark{1,2}, Defang Kong\altaffilmark{1,2}, Zhike Xue\altaffilmark{1,2}, Liheng Yang\altaffilmark{1,2}, Qiaoling Li\altaffilmark{1,4}}

\altaffiltext{1}{Yunnan Observatories, Chinese Academy of Sciences, Kunming 650011, People.s Republic of China.}
\altaffiltext{2}{Center for Astronomical Mega-Science, Chinese Academy of Sciences, 20A Datun Road, Chaoyang District, Beijing, 100012, People.s Republic of China}
\altaffiltext{3}{CAS Key Laboratory of Solar Activity, National Astronomical Observatories, Beijing 100012, China}
\altaffiltext{4}{University of Chinese Academy of Sciences, Yuquan Road, Shijingshan Block Beijing 100049, People.s Republic of China}
\begin{abstract}
With the launch of SDO, many EUV waves were observed during solar eruptions. However, the joint observations of Moreton and EUV waves are still relatively rare. We present an event that a small-scale filament eruption simultaneously results in a Moreton wave, an EUV wave and a Coronal Mass Ejection in active region NOAA 12740. Firstly, we find that some dark elongate lanes or filamentary structures in the photosphere existed under the small-scale filament and drifted downward, which manifests that the small-scale filament was emerging and lifting from subsurface. Secondly, combining the simultaneous observations in different Extreme UltraViolet (EUV) and H$\alpha$ passbands, we study the kinematic characteristics of the Moreton and EUV waves. The comparable propagating velocities and the similar morphology of Moreton and different passbands EUV wavefronts were obtained. We deduce that Moreton and different passbands EUV waves were the perturbations in different temperature-associated layers induced by the coronal magneto-hydrodynamic shock wave. We also find the refracted, reflected and diffracted phenomena during the propagation of the EUV wave. By using power-law fittings, the kinematic characteristics of unaffected, refracted and diffracted waves were obtained. The extrapolation field derived by the potential field source surface (PFSS) model manifests that the existence of an interface of different magnetic system (magnetic separatrix) result in refraction, reflection and deviation of the EUV wave.
\end{abstract}


\section{Introduction} \label{sec:intro}
As is well known, coronal mass ejections (CMEs) and flares are the most spectacular manifestations of solar activity, which result from rapid change in the coronal magnetic field. They have a close relationship with the filament eruptions and coronal jets (e.g., \citealp{mun79,zha01,alo10,zhe16,zhe17,wan17,yan17}). Either filament eruptions or coronal jets are generally considered to be the initial sources of these energetic events \citep{rao16,hon17}. Many observations and numerical computations/simulations show that magnetic reconnection plays a key role in these energetic releasing processes of the flares and CMEs (e.g., \citealp{par63,low96,lin00,mei12a,xue16,yan18}). On the other hand, these violently energetic events are often accompanied with many additional solar activities (e.g., rearrangements of magnetic fields, solar energetic particles, shocks, radio bursts, and chromospheric or coronal perturbations like Moreton or EUV waves), which show a significant influence on the solar atmosphere \citep{koz11,she18,xu19}.

Moreton waves are flare-associated phenomena, which were first reported about 60 years ago \citep{mor60,more60}, propagating in restricted angles with speeds of 500-1500 km s$^{-1}$ and observed in chromospheric spectral lines (typically in H$\alpha$ and He I 10830 $\rm\AA$) \citep{ath61,vrs02,war04,bal10,liu13}. They often appear as arc-shaped fronts propagating away from the flaring active region that last for only several minutes and become irregular and diffuse. In the remote region, they often cause the activation or ``winking" of filaments \citep{tri09,asa12}. \citet{uch68} explained them as the enhancing signatures in chromospheric spectral line's emission when the coronal wavefront of the shock produced by a coronal fast-mode shock wave sweeps the chromosphere. Uchida's interpretation has been widely accepted, but the coronal counterpart of Moreton wave is under debate for a few decades. Especially, it is rarely observed the Moreton waves even in the large flares.

Another wave-like perturbations associated with solar flares in the corona, coronal bright fronts (CBFs), were first discovered in EUV band images by using the observation from the Extreme-ultraviolet Imaging Telescope (EIT; \citealp{del95}) onboard the Solar and Heliospheric Observatory (SOHO; \citealp{dom95}), and also were named EIT waves \citep{mos97,tho98}, which were expected to be the coronal counterpart of the Moreton waves. Unfortunately, they show different physical characteristics from the Moreton waves. They propagate across solar surface at velocity of about 200-400 km s$^{-1}$ which is much slower than that of the Moreton wave. Their lifetime is much longer than that of Moreton wave which is about 45-60 minutes \citep{kla00,tho09}. Otherwise, they are also found to experience reflection and refraction, while they have been observed to remain stationary at the boundary of coronal holes \citep{tho98,tri07,xue13}. Owing to these discrepancies between the EIT waves and Moreton waves, it is hard to evidence that the EIT waves are the coronal counterparts of Moreton waves. On the other hand, the nature of EIT waves is also still under debate. Different models have been proposed to give an interpretation of EIT waves, which include fast-mode magnetoacoustic wave model \citep{wan00,ofm02,bal05,ver08,sch10}, slow-mode wave model \citep{wil06,wan09}, shock echo model \citep{mei12,xie19}, and pseudo wave/field-line stretching model \citep{del99,che02}. According to the field-line stretching model, EIT bright fronts are not the real wave, but the plasma compression due to successive stretching or opening of closed field lines driven by an erupting flux rope. This model can resolve inconsistency of the propagating velocity between EIT and Moreton waves, which also can answer the question of why EIT waves often stop at magnetic separatrices.

In high temporal and spatial resolution observations, it is found that more than one type of EUV waves exist simultaneously during some erupting events. Slow and fast EUV waves have been reported in some investigations \citep{har03,liu10,che11,she12}. The slow EUV waves are thought to be corresponding to historical EIT waves. \cite{che11} found that a fast moving wave front with a speed of 560 km s$^{-1}$ existed ahead of an slow wave with a speed of $\sim$ 190 km s$^{-1}$ in a microflare event on 2010 July 27. The fast EUV waves have been studied significantly since the launch of Solar Dynamic Observatory (SDO), which are thought to be shock waves associated with chromospheric Moreton waves and fast-MHD shock waves \citep{liu10,koz11,ma11,asa12,fra16}. Their velocities could be up to $\sim$ 1000 km s$^{-1}$ \citep{koz11}. Otherwise, they are bright fronts visible in the EUV difference images except for 171 $\rm\AA$ images, which exhibit dark wavefronts in 171 $\rm\AA$ images \citep{liu10,ma11}. Despite decades of research, it is quite rare to jointly capture the EUV and Moreton waves. Therefore, it still need more evidences to confirm whether the fast EUV waves are the counterpart of Moreton wave in the corona and determine the relationship between EUV and Moreton waves.

On the other hand, the excitation of these coronal shock waves is another subject of long-lasting study and debate. Two popular views on the origin of these shock waves, have been proposed. One view is that CMEs associated with filament eruption are considered as the driver of these shock waves \citep{bie02,gre11,bal10,ura19}. In this view, the coronal shock wave is considered to be generated by the combination of the projectile effect and 3D piston, as the CME moves through the ambient plasma and expands its body in all directions. The other view suggests that the shock waves are generated by a pressure pulse in the flare region \citep{vrs08,tem09,cli16}. In this view, the shock is created in a temporary piston generated by the source region expansion caused by the flare energy release. \cite{vrs06} identified a second coronal wave behind the CME structure and concluded that the wave was not driven or launched by the CME, but rather by the flare. However, \cite{bie02} shown a closely correlation between EUV waves and CMEs and a significantly weak correlation between EUV waves and flares by a large statistic studies of EUV wave observations. In the recent years, a lot of evidences have been presented that favors CMEs over flares as the cause of coronal waves (e.g., \citealp{hud03,ma11}). Still, the possibility that some waves are caused by flare-associated pressure pulses cannot be ruled out. Up to now, the origin of these shock waves is yet an open question for understanding these large-scale globally perturbations in the solar atmosphere.

In this paper, we will present the simultaneous observations of Moreton and EUV waves associated with a small-scale filament eruption, and analyze the morphology and the kinematics of the wave. Furthermore, we also study the characteristics of the small-scale filament before the eruption. The sections of this paper are organized as follows: the observations and methods are described in Section \ref{sec:obser methods}, the results are presented in the Section \ref{sec:results}, and the summary and discussions are given in Section \ref{sec:conclusion}.

\section{Observations and Data Analysis} \label{sec:obser methods}
The data set are primarily from New Vacuum Solar Telescope\footnote{\url{http://fso.ynao.ac.cn}} (NVST; \citealp{liu14,yan19}), Atmospheric Imaging Assembly (AIA; \citealp{lem12}) onboard the SDO\footnote{\url{https://sdo.gsfc.nasa.gov}} \citep{pes12} and Global Oscillation
Network Group (GONG) instruments\footnote{\url{https://gong.nso.edu}} \citep{har96,har11} . The NVST is a vacuum solar telescope with a 985 mm clear aperture, located at Fuxian Lake, in Yunnan Province, China. It provides the simultaneously high-resolution imaging observations of photosphere (TiO 7058 $\rm\AA$) and chromosphere (H$\alpha$ 6562.8 $\rm\AA$) to study the small-scale filament before the eruption. The TiO images have a pixel size of 0.\arcsec052 and a cadence of 30 s. The H$\alpha$ images include three passbands: H$\alpha$ blue-wing (-0.4 $\rm\AA$), H$\alpha$ center, and H$\alpha$ red-wing (+0.4 $\rm\AA$), with a 45 s cadence and a spatial resolution of 0.\arcsec165 per pixel. The field of view (FOV) of H$\alpha$ images is 150\arcsec $\times$ 150$\arcsec$ while that of TiO images is 120\arcsec $\times$ 100\arcsec. These data are calibrated from Level 0 to Level 1 with dark current subtracted and flat field corrected, and then speckle masking method was used to reconstruct the calibrated images from Level 1 to Level 1+ by the method described in \citet{xia16}.

The AIA instrument on SDO can provide seven simultaneous full-disk EUV images of the low corona and transition region with a pixel spatial size of 0.\arcsec6 and a cadence of 12 s, which includes the channels centered at 94 $\rm\AA$, 131 $\rm\AA$, 171 $\rm\AA$, 193 $\rm\AA$, 211 $\rm\AA$, 304 $\rm\AA$ and 335 $\rm\AA$. We mainly use the channels of 211 $\rm\AA$, 193 $\rm\AA$, 171 $\rm\AA$, 304 $\rm\AA$ to study the characteristics of EUV wave. The 211 $\rm\AA$ and 193 $\rm\AA$ images are dominated by the Fe XIV lines (logT$\sim$6.3) and Fe XII lines (logT$\sim$6.2) for active-region observations, respectively. The 171 $\rm\AA$ and 304 $\rm\AA$ images are dominated by the Fe IX line (log T$\sim$5.85) and He II line (log T$\sim$4.7) \citep{lem12}. Full-disk H$\rm\alpha$ images from six GONG stations are used to study the Moreton wave. The spatial resolution of the H$\rm\alpha$ images is about 1$\arcsec$ per pixel and the cadence is 1 minute. Both of AIA/SDO EUV and GONG H$\alpha$ images were de-rotated to a reference time (04:59:04 UT) to trace the interesting region due to the the Sun rotation and correct for the displacement of coronal structures caused by the differential rotation of the Sun. Otherwise, line-of-sight (LOS) magnetic fields from the Helioseismic and Magnetic Imager (HMI; \citealp{sch12}) onboard SDO with a pixel spatial size of 0.\arcsec5 and a cadence of 45 s, continuum intensity images from Large Angle Spectroscopic Coronagraph onboard SOHO (LASCO/SOHO) C2 \citep{bru95} and soft X-ray 1-8 $\rm\AA$ flux data from Geostationary Operational Environmental Satellite (GOES) are also utilized in this study.
\section{Results} \label{sec:results}
\subsection{a small-scale filament}
\begin{figure}[h!]
\figurenum{1}
\plotone{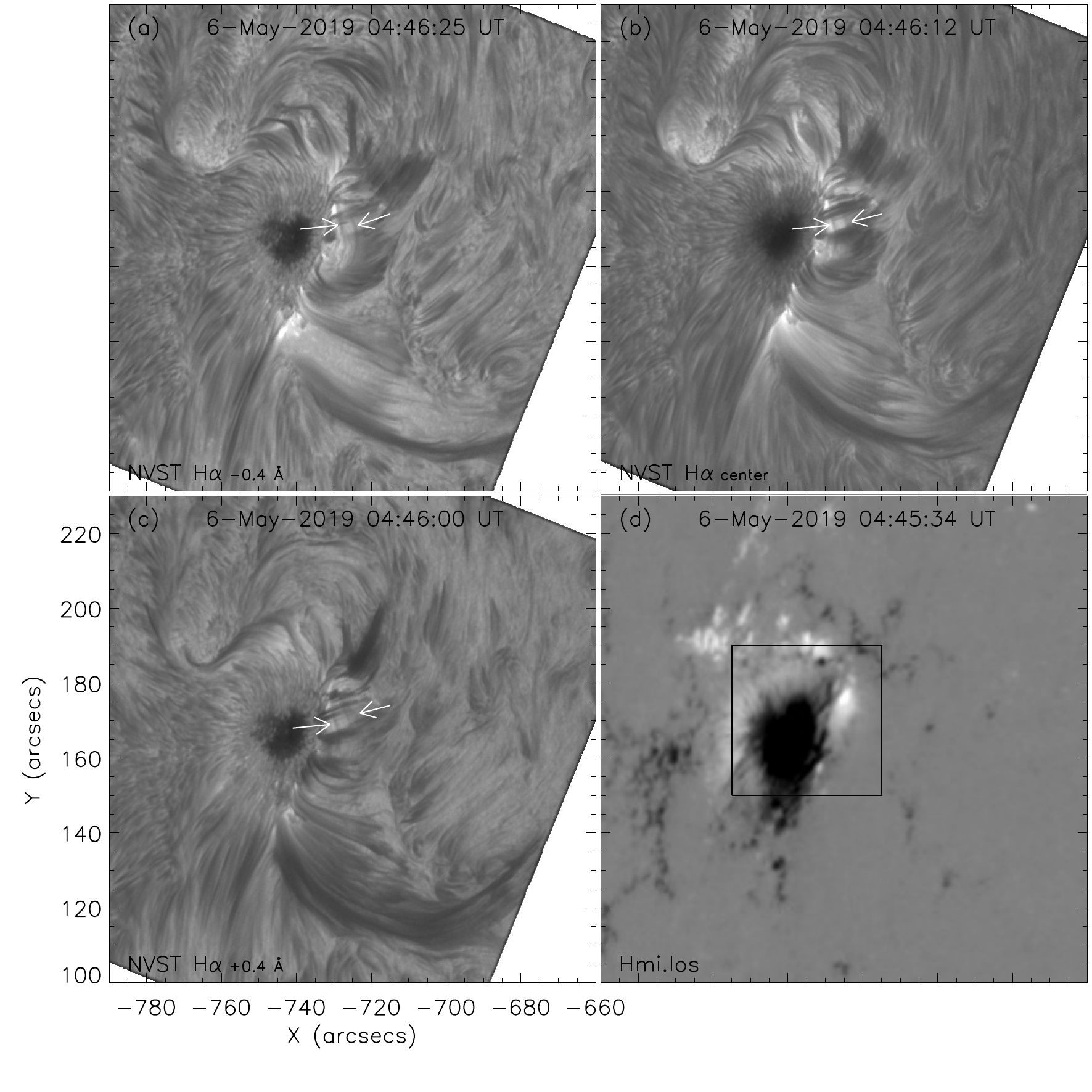}
\caption{The small-scale filament observed by NVST H$\alpha$ before the eruption. (a) H$\alpha$ -0.4 $\rm\AA$ image. (b) H$\alpha$ center image. (c) H$\alpha$ +0.4 $\rm\AA$ image. (d) Line of sight magnetic field observed by SDO/HMI. White arrows in panels (a)-(c) indicate the filamentary structures of the small-scale filament. The field of view of Fig.\ref{fig2} is marked by the black box in the panel (d).}\label{fig1}
\end{figure}

On 2019 May 06, Active Region NOAA 12740 with $\rm\beta \delta/ \alpha$ magnetic configuration appeared at the northeast of the Sun (N08E41), which comprised of a leading sunspot with negative polarities and complex diffuse following positive polarities. Figs.\ref{fig1} (a)-(c) show the images of H$\alpha$ -0.4 $\rm\AA$, H$\alpha$ center, H$\alpha$ +0.4 $\rm\AA$ observed by the NVST at around 04:46 UT before the filament eruption, while panel (d) exhibits the corresponding LOS magnetic field observed by HMI/SDO. A small-scale filament could be discerned in the vicinity of west side of the main leading sunspot (see in panels (a)-(c)), which resided under the flickering filamentary arcade. The small-scale filament consisted of two filamentary structures marked by two white arrows in each panels (a)-(c). Owing to the location of the region of interest close to the solar limb, the LOS magnetic field would be significantly influenced by the projection effect. Nevertheless, we also could distinguish that the northern footpoint of filament inhabited at positive polarity while the southern footpoint rooted in the vicinity of the negative polarity nearby the main sunspot.

\begin{figure}[h!]
\figurenum{2}
\plotone{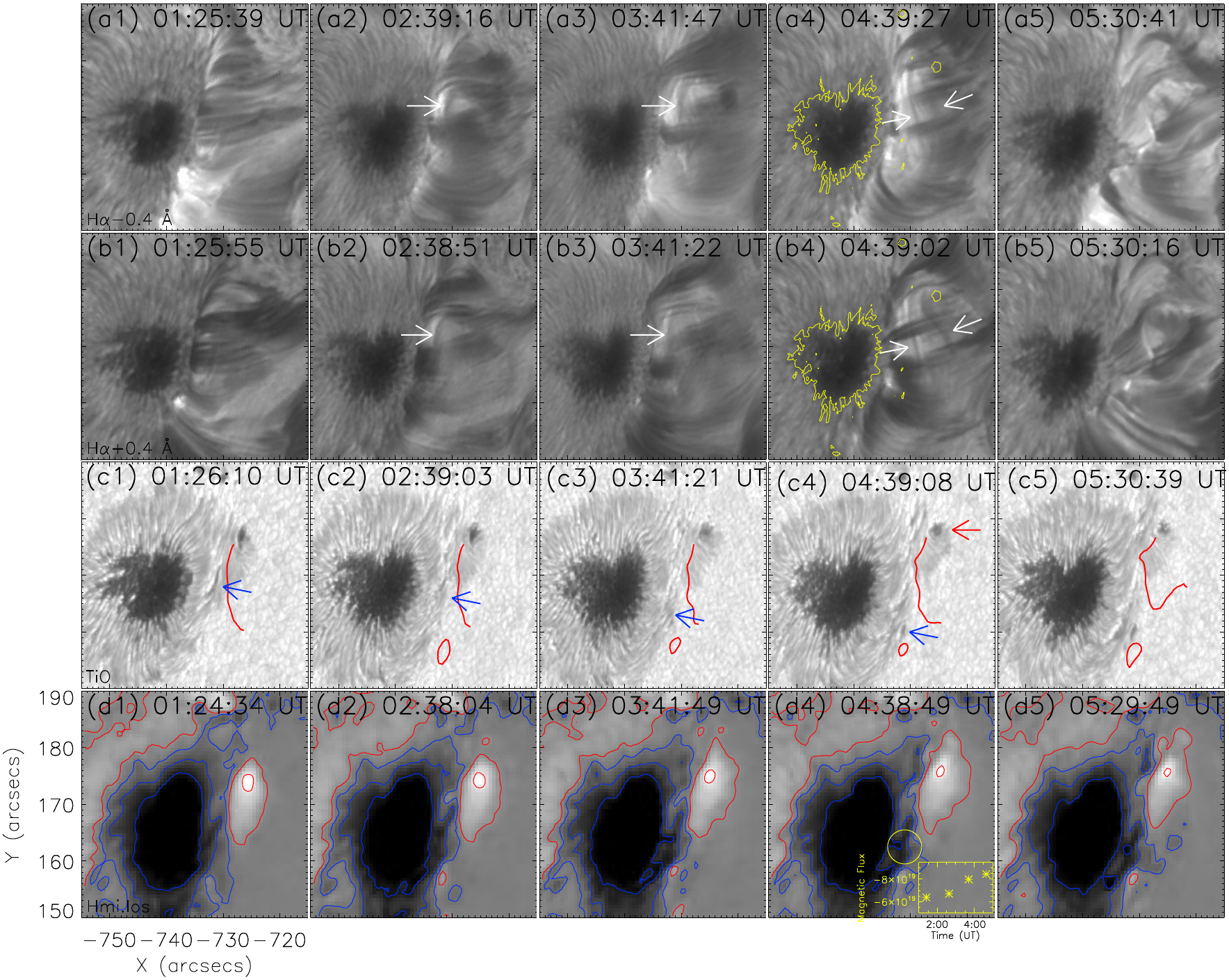}
\caption{Evolution of the small-scale filament before the eruption. (a1)-(a5) H$\alpha$ blue wing images (H$\alpha$ -0.4 $\rm\AA$). (b1)-(b5) H$\alpha$ red wing images (H$\alpha$ +0.4 $\rm\AA$). (c1)-(c5) TiO images observed by NVST. Red lines denote the PILs in the photosphere. (d1)-(d5) Line of sight magnetic field from SDO/HMI. The green contours in the panels (a4) and (b4) is the contour of corresponding TiO image at 04:39:08 UT. The red and blue contours in the panel (d1) are the magnetic field with the levels of 800, 100 and -100, -400, -800 G, respectively. The profile of red asterisks the panel (d4) denotes the variation of negative magnetic flux inside the yellow circle.}\label{fig2}
\end{figure}

Fig.\ref{fig2} exhibits the evolution of the small-scale filament during the period from around 01:25 UT to 05:30 UT. Panels (a1)-(a5) and (b1)-(b5) are the low chromospheric observations in H$\alpha$ -0.4 $\rm\AA$ and H$\alpha$ +0.4 $\rm\AA$, respectively. Panels (c1)-(c5) and (d1)-(d5) are the corresponding photospheric observations of the TiO images from NVST and LOS magnetograms from HMI/SDO, respectively. At around 01:25 UT, the filament was difficult to distinguish due to the thick overlying filamentary arcade or the filament inhabited too low to be observed in the chromosphere. The interesting phenomenon was found that an elongate dark lane had existed nearby the polarity inversion line (PIL) (red lines in the panels (c1)-(c5)) in the photosphere which was perpendicular to the penumbral fibers instead of parallel to it (see panel (c1)). It should notice that the PIL remained significant uncertainty due to the projection effect in magnetic field. This type elongate filamentary structure in general is representation of the strong transverse magnetic field \citep{ots07,zha18,wan18}. Therefore, we deduce that a flux tube with strong transverse magnetic field lying on the position marked by the blue arrow in the panel (c1). At around 02:39 UT, the small-scale filament could be distinguished from the images of H$\rm\alpha$ blue and red wings (see panels (a2)\&(b2)), which was marked by the white arrow in each panel. At around 03:42 UT, the small-scale filament became more and more visible under the overlying arcade structures, indicated by white arrows in panels (a3)\&(b3). At the meanwhile, the dark lane in the photosphere moved toward solar south and became lighter, which was marked by the blue arrow in the panel (c3). At around 04:39 UT, the filament split into two elongate filamentary structures marked by two white arrows in panels (a4)\&(b4), while the dark lane on the photosphere kept on moving down. On the other hand, the dark lane always located in the northern of some small closed PILs (see panel (c2)-(c4)). This means the dark lane was not associated with the emergence of an other new dipole nearby some small closed PILs. Based on the contours of TiO overlaid on the H$\alpha$ images (see panels (a4)\&(b4)), we deduced that the filament eventually rooted around the pore and the dark lane marked by the red and blue arrows in the panel (c4), respectively. Comparing the positions of the pore and the dark lane, the filament's footpoints separated more and more farther away from each other before the filament eruption. After its eruption, this small-scale filament was completely ejected and it disappeared at around 05:30 UT (see panels (a5) \& (b5)). The corresponding LOS magnetic fields are shown in the panels (d1)-(d5). According to the profile of the asterisks in bottom right of panel (d4), the negative flux inside the yellow circle shown an increasing tendency. This means that the negative flux generally enhanced at the southern footpoint of the filament. This signal indicated that flux emergence occurred in this place. Based on the separation of the filament footpoint and the flux emergence in the footpoint, we speculate that the small-scale filament had experienced emerging and lifting process. At the beginning, only a part of the flux tube exposed and lain in the photosphere. The strong transverse magnetic field in the flux tube was observed as the dark lane in the photosphere. As the flux tube lift up, the dark lane became lighter and moved toward south. When the top of flux tube entirely departed from the photosphere, the dark lane would represent the footpoint of the flux tube with strong transverse magnetic field in the photosphere. As the flux tube kept lifting up, the small-scale filament could be visible in the chromosphere and the footpoints of the filament would move away from each other. Due to the emergence of the flux tube, magnetic flux in the photosphere also increased during this process, which caused the enhancement of the magnetic flux at the footpoint of the filament marked by the yellow circle in the panel (d4).

\subsection{the eruption of the small-scale filament}
As the small-scale filament rose up, the filament became more visible and unstable. And then, the small-scale filament start to erupt at around 05:05 UT (see the animation of Fig.\ref{fig3}). It triggered a M 1.0 flare and a narrow CME. Fig.\ref{fig3} (a) shows the image of EUV 193 $\rm\AA$ at 05:05:28 UT. Some brightenings marked by the red arrow were found in the vicinity of the filament. The blue curve in upper right of the panel (a) is the variations of the GOES soft X-ray 1-8 $\rm\AA$ flux. Three vertical dashed red lines denote the onset, peak and end of the flare, respectively. The flare started at 05:05 UT, peaked at 05:10 UT and ended at 05:16 UT. This flare lasted exceedingly short time, only about 11 minutes. It should be a impulse-type or compact flare. This is related to processes of the filament eruption. {Panel (b) shows the flare in 131 $\rm\AA$ at peak time of 05:10 UT. We use the intensity-weight of the flare region at this time to determine the eruption center. The determined eruption center (x=-728$\arcsec$, y=174$\arcsec$) is marked by the blue asterisk in the panel (b). Panels (c1)-(c4) and (d1)-(d4) show observations at different moments in H$\alpha$ blue-wing and H$\alpha$ center, respectively. Before the eruption, the filament resided under the overlying filamentary arcades and had turned to more bending (see panels (c1), (c2)\&(d1)). Several brightenings had occurred nearby the north part of the filament, which were marked by the red circles in the panels (c2) \& (d2). During the eruption, the active filament lifted and interacted with the overlying arcades. And then the filament was collapsed and the materials were ejected along the magnetic field. After the eruption, the small-scale filament was absent (see panels (c4)\&(c4)). The more details can be seen from the animation of Fig.\ref{fig3}. The eruption process would be alike the mechanism of the blowout-type jet \citep{moo10}. Fig.\ref{fig4} shows the evolution of the CME. At around 05:36 UT after about 30 minutes of the eruption, a narrow CME with angle width 52 degree was observed by LASCO/SOHO C2. The mean velocity of the CME was about 376 km s$^{-1}$. An interesting feature was found that some dark flowing structure experienced shrinking motion after the bright CME front ejected out (see panels (c)\&(d)). The material flow in the corona is along the magnetic field. Therefore, it means that the magnetic field should also experience shrinking motion. More details was exhibited in the animation of Fig.\ref{fig4}. The impulse-type flare and shrinking narrow CME may also manifested that the filament erupted as the blowout-type jets \citep{dua19}.

\begin{figure}[t!]
\figurenum{3}
\plotone{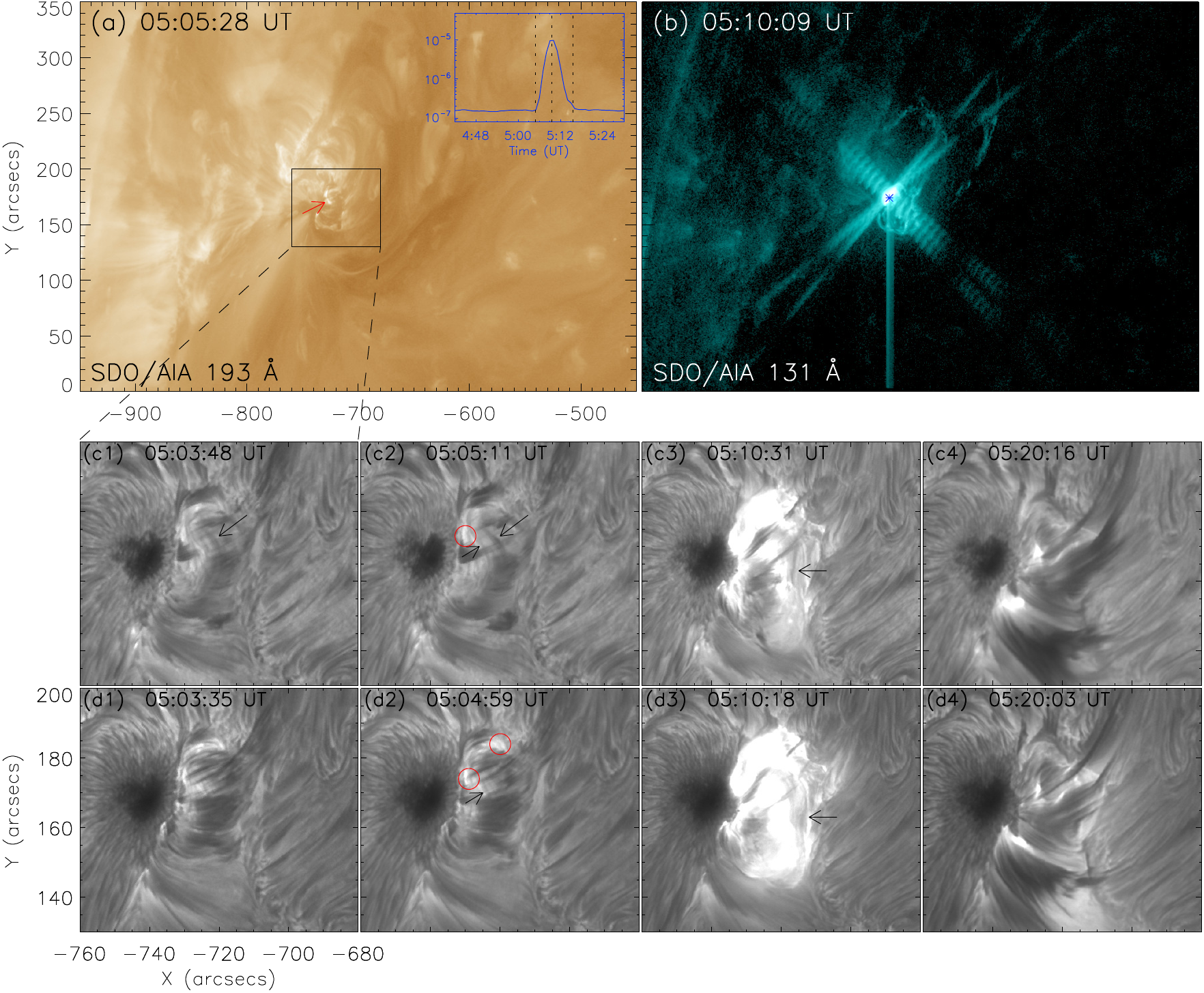}
\caption{The eruption process of the small-scale filament. (a) SDO/AIA 193 $\rm\AA$ image at 05:05:28 UT. The blue curve in panel (a) depict the variation of soft X ray 1-8 $\rm\AA$ fluxes from GOES. Three vertical dashed red lines indicate the moments of onset, peak and end of the flare. (b) SDO/AIA 131 $\rm\AA$ image at 05:10:09 UT. The blue asterisk marks the eruption center. (c1)-(c4) H$\alpha$ -0.4 $\rm\AA$ images observed by NVST. (d1)-(d4) H$\alpha$ center images observed by NVST. The black arrows mark the filament structures.}\label{fig3}
\end{figure}
\begin{figure}[t!]
\figurenum{4}
\plotone{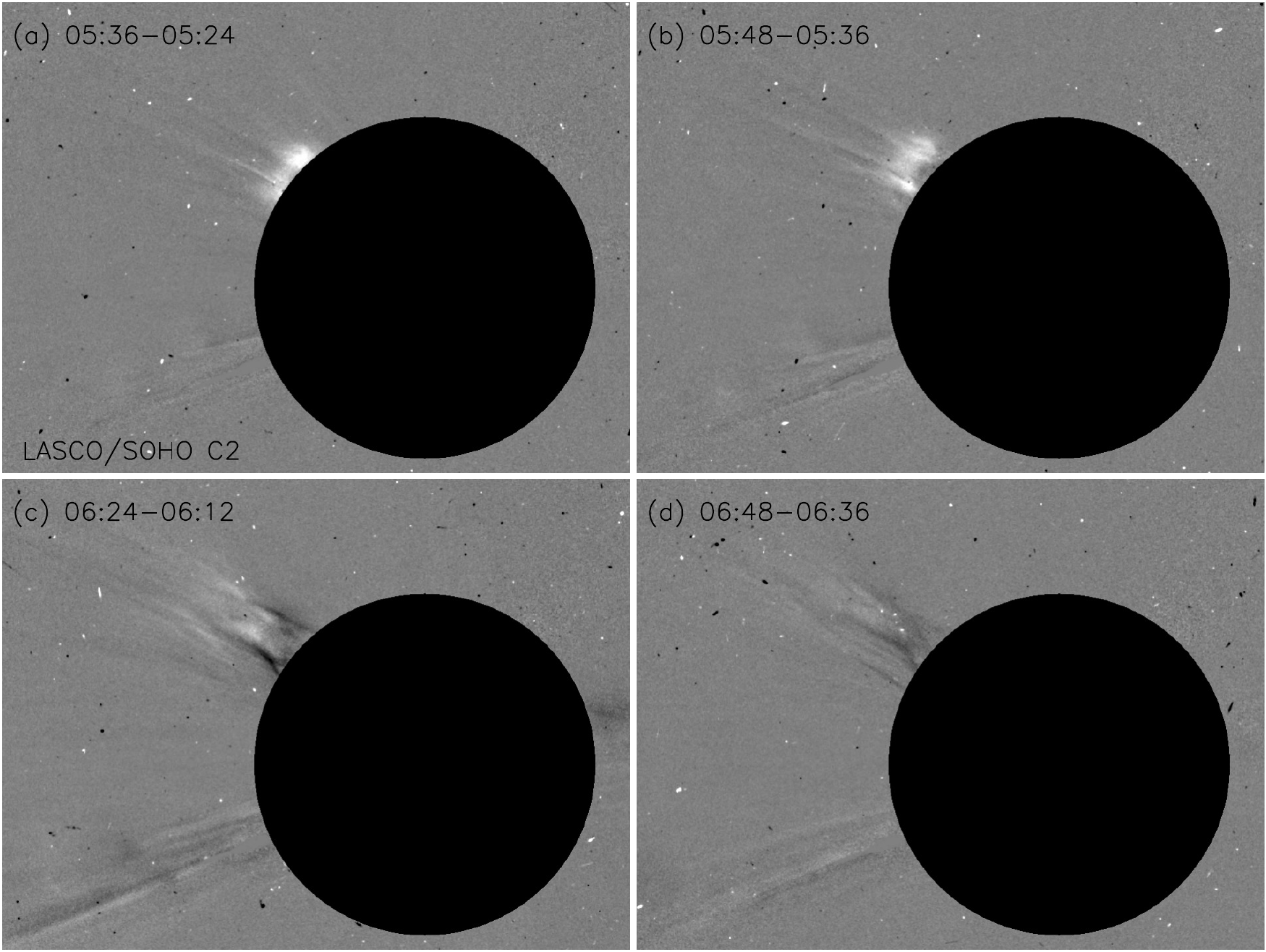}
\caption{The evolution of the CME observed by LASCO/SOHO C2. (a)-(d) Running-difference images of continuum intensity image.}\label{fig4}
\end{figure}
\subsection{Moreton wave and EUV wave}

As the small-scale filament erupted, same accompanied global perturbations including Moreton and EUV waves could be discerned (see Figs.\ref{fig5} \& \ref{fig7}). Fig.\ref{fig5} shows the running-difference images of H$\alpha$ from GONG in different moments. Due to 1 minute cadence of GONG observation, the running-difference image of GONG was derived from the current image subtracting the image 1 minute earlier. From Fig.\ref{fig5} and the animation of Fig.\ref{fig5}, an arc-shaped Moreton wave could be distinguished, which originated from the flare site and propagated towards solar north on the chromospheric disk. The blue dashed curve line in the panel (d) depicts the wavefront of the Moreton wave at 05:10:50 UT. We use the intensity profiles technique \citep{liu10,fra16} to determine the kinematic properties of the Moreton wave, which new ``north pole" is set at the eruption center (x=-728$\arcsec$, y=174$\arcsec$). The intensity profile of the sector was derived by averaging intensities of pixels in the ``corresponding latitudinal" direction, which was a function of distance measured from the eruption center along the ``corresponding longitudinal" circle. The measurement of the distance in this technique is corrected by the sphericity of the solar atmosphere as the method in \cite{liu10}. The sector is drawn by the white solid line in the panel (a) of Fig.\ref{fig6}. The angle extent of the sector is 25$\degr$. Figs.\ref{fig6} (c)-(f) present the intensity profiles along the sector derived by the H$\alpha$ running-difference image in different moments. The propagation of the Moreton wave could be identified by peak and trough signals, which marked by the blue and red arrows, respectively. We use three proxies to derive the velocity of the Moreton wave: onset of increase, peak (marked by blue arrows), minimum of the trough (marked by red arrows). The black asterisks, blue diamonds and red triangles in Fig.\ref{fig6} (b) indicate the distances of these three proxies as a function of time, respectively. On the other hand, the linear fitting has been implemented on these profiles of proxies. The dotted lines of different colors in the panel (b) are the fitted lines of fitting different proxies. According to the slope of these fitted lines, the velocities of three proxies can be derived, which are about 574.8 $\pm$ 20.6 km s$^{-1}$, 625.2 $\pm$ 41.2 km s$^{-1}$ and 635.3 $\pm$ 61.7 km s$^{-1}$, respectively. The errors are estimated as the standard error of fitting divided by time. Therefore, the average velocity of the Moreton wave in our study have been estimated to be about 611.8 $\pm$ 41.2 km s$^{-1}$.
\begin{figure}[t!]
\figurenum{5}
\plotone{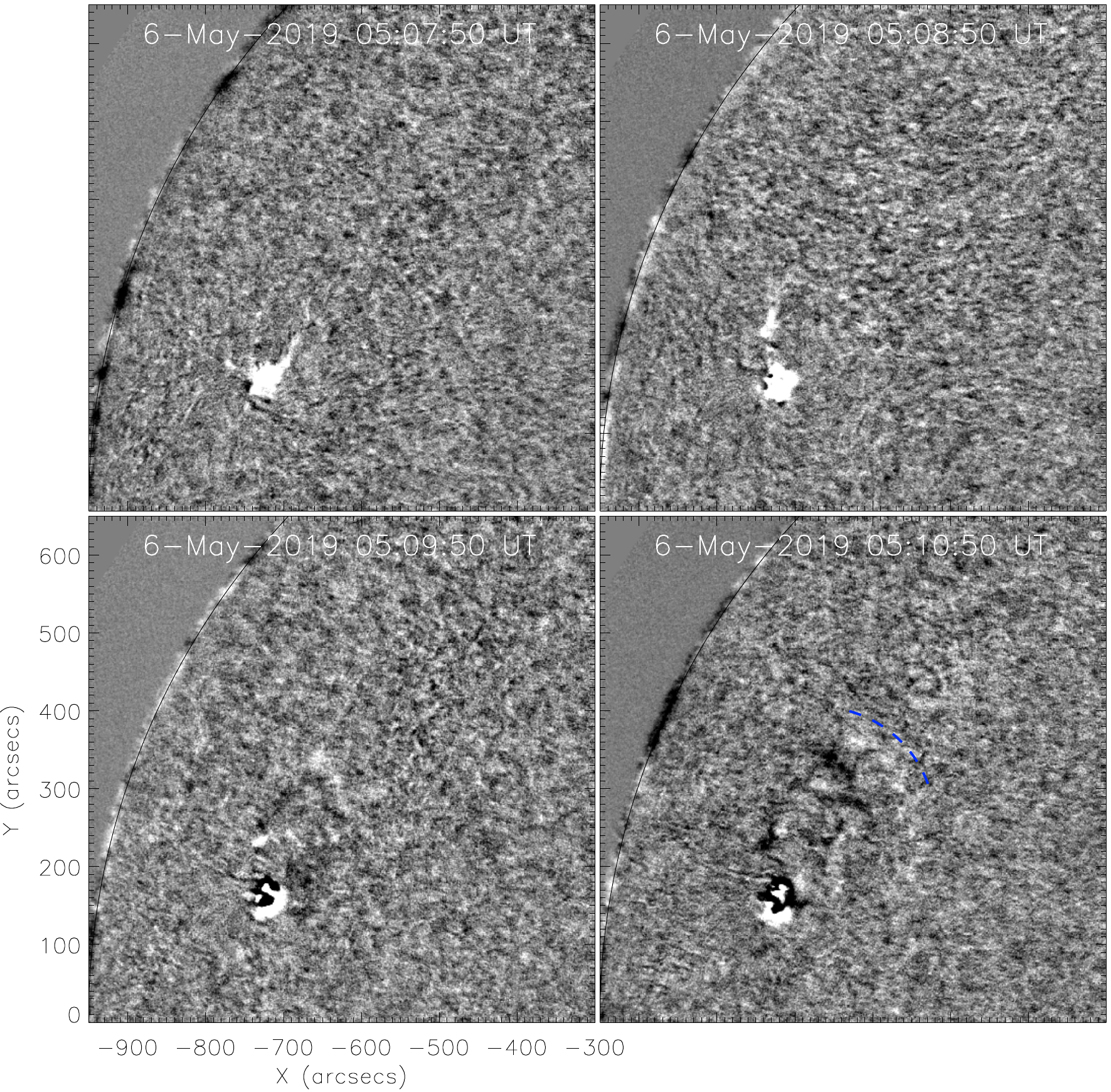}
\caption{Evolution of Moreton wave observed by GONG. (a)-(d) H$\alpha$ running difference images. The black curves in different panels indicate the solar limb. The blue dashed arc-shaped line in the panel (d) marks the wavefront of the Moreton wave at 05:10:50 UT. A animation of GONG H$\alpha$ running difference image is available.}\label{fig5}
\end{figure}

\begin{figure}[t!]
\figurenum{6}
\plotone{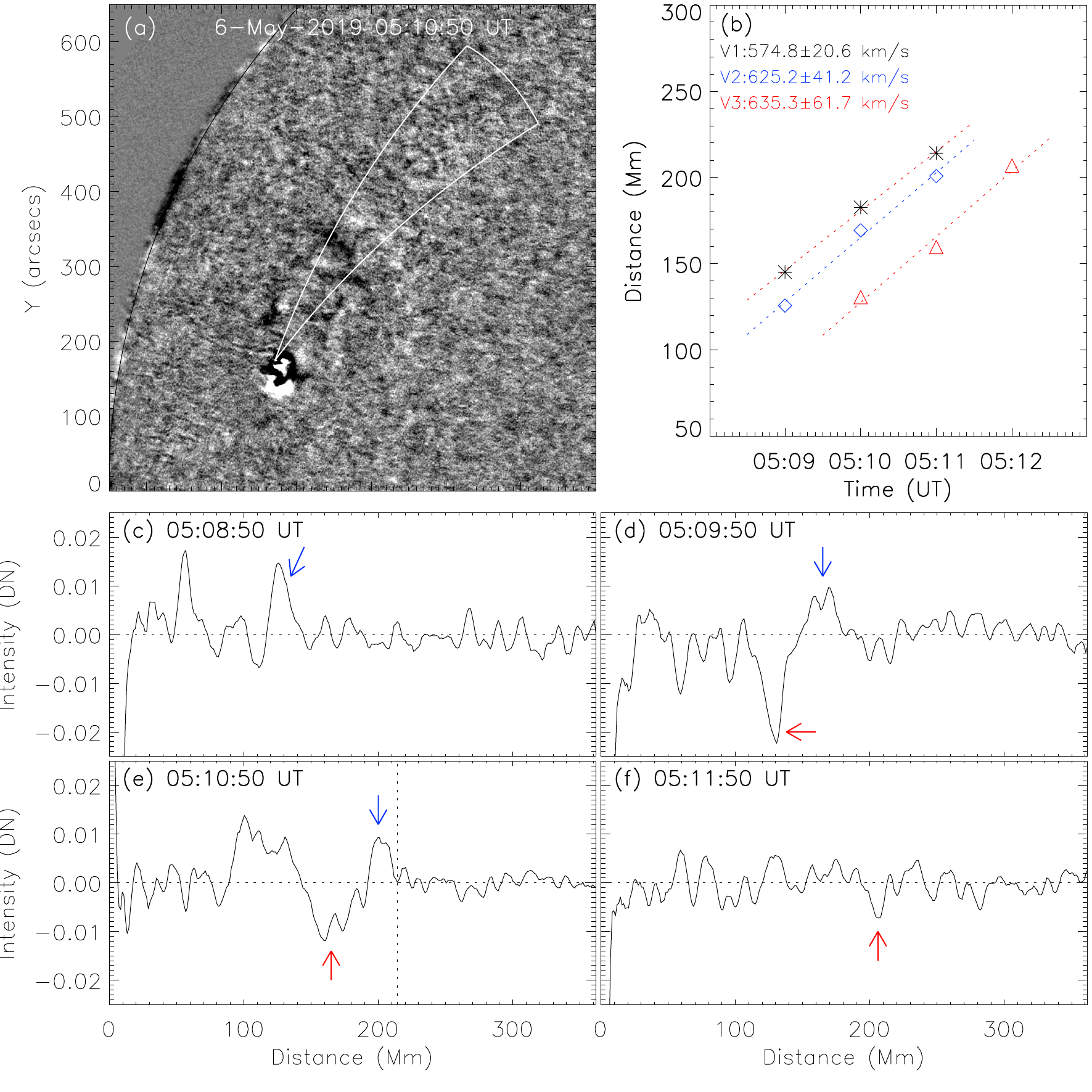}
\caption{(a) H$\alpha$ running difference image at 05:10:50 UT. The white fan-shape denotes the sector for Moreton wave. (b) The distances of three proxies from the eruption center as the function of time. The black asterisks, blue diamonds and red triangles are represent the onset of increase, peak, minimun of the trough in the intensity profiles of the sector related to the Moreton wave, respectively. (c)-(f) The intensity profiles of the sector in different moments. The blue and red arrows mark the different signatures of the Moreton wave. The vertical dotted line in panel (e) denotes the wavefront position at 05:10:50 UT.}\label{fig6}
\end{figure}

In the meanwhile, EUV wave also can be identified by using SDO/AIA different passbands images. Fig.\ref{fig7} shows the EUV waves in running difference images of 193 $\rm\AA$, 211 $\rm\AA$, 304 $\rm\AA$ and 171 $\rm\AA$ at around 05:11 UT. The EUV wavefront in 193 $\rm\AA$, 211 $\rm\AA$ and 304 $\rm\AA$ passbands is bright, while the one in 171 $\rm\AA$ is dark. This mean that the EUV wavefront enhanced the radiative intensity of 193 $\rm\AA$, 211 $\rm\AA$ and 304 $\rm\AA$ but reduced the radiative intensity of 171 $\rm\AA$. This feature was also found by previous studies \citep{liu10,ma11,li12}. This feature suggests that the EUV wavefront has heated the iron elements above the characteristic temperature of the 171 $\rm\AA$ passband. The 304 $\rm\AA$ signature is faint while the signatures of 193 $\rm\AA$ and 211 $\rm\AA$ are relatively distinct. We overlay the wavefront of Moreton wave on the EUV wave images at almost the same moment, which is presented by the blue dashed curve in each panel of Fig.\ref{fig7}. It is found that the EUV bright wavefront in 304 $\rm\AA$ showed a similar aspect with the Moreton wavefront. However, EUV bright wavefronts in 193 $\rm\AA$, 211 $\rm\AA$ and the dark wavefront in 171 $\rm\AA$ were in the front of the Moreton wavefront, which is similar to the result presented by \cite{vrs02}. The yellow curve in the bottom of each panel exhibits the variations of intensity of different passbands along the sector. The wavefront distances (D) from the eruption center in different passbands could be derived by using these curves of \textbf{running-difference intensity}. The distance where the curve or its derivative equals to zero nearby the wavefront, represents the wavefront distance, which is marked by the vertical line in each plot. We obtain that D$_{193}$(237 Mm) $\sim$ D$_{211}$(236 Mm) $>$ D$_{171}$(221 Mm) $>$ D$_{304}$(218 Mm), while D$_{Moreton}$ is about 214 (Mm) at 05:10:50 (see the panel (e) of Fig.\ref{fig6}). Combining the analysis in the Appendix, when the wave almost propagate along the path parallel to the solar limb, the error caused by the measurement is tiny and can be negligible. As is well known, different passbands images reveal the information in different solar layer with different temperature. Based on the demonstration in the Section 2, the dominating temperature responses in different EUV passbands is that: logT$_{211}(\sim6.3)$ $\sim$ logT$_{193}(\sim 6.2)$ $>$ logT$_{171}(\sim 5.85)$ $>$ logT$_{304}(\sim 4.7)$. In other words, the height of the wavefronts observed by 211 $\rm\AA$, 193 $\rm\AA$ and 171 $\rm\AA$ are higher than that of the wavefront observed by 304 $\rm\AA$ and Moreton wavefront. It is reasonable to consider that the wavefront in different passbands just show different aspects of one shock wave in different solar atmosphere layers and the shock wave was slant in the solar height direction. Therefore, the EUV wave and Moreton wave would be caused by the same shock wave.

\begin{figure}[t!]
\figurenum{7}
\plotone{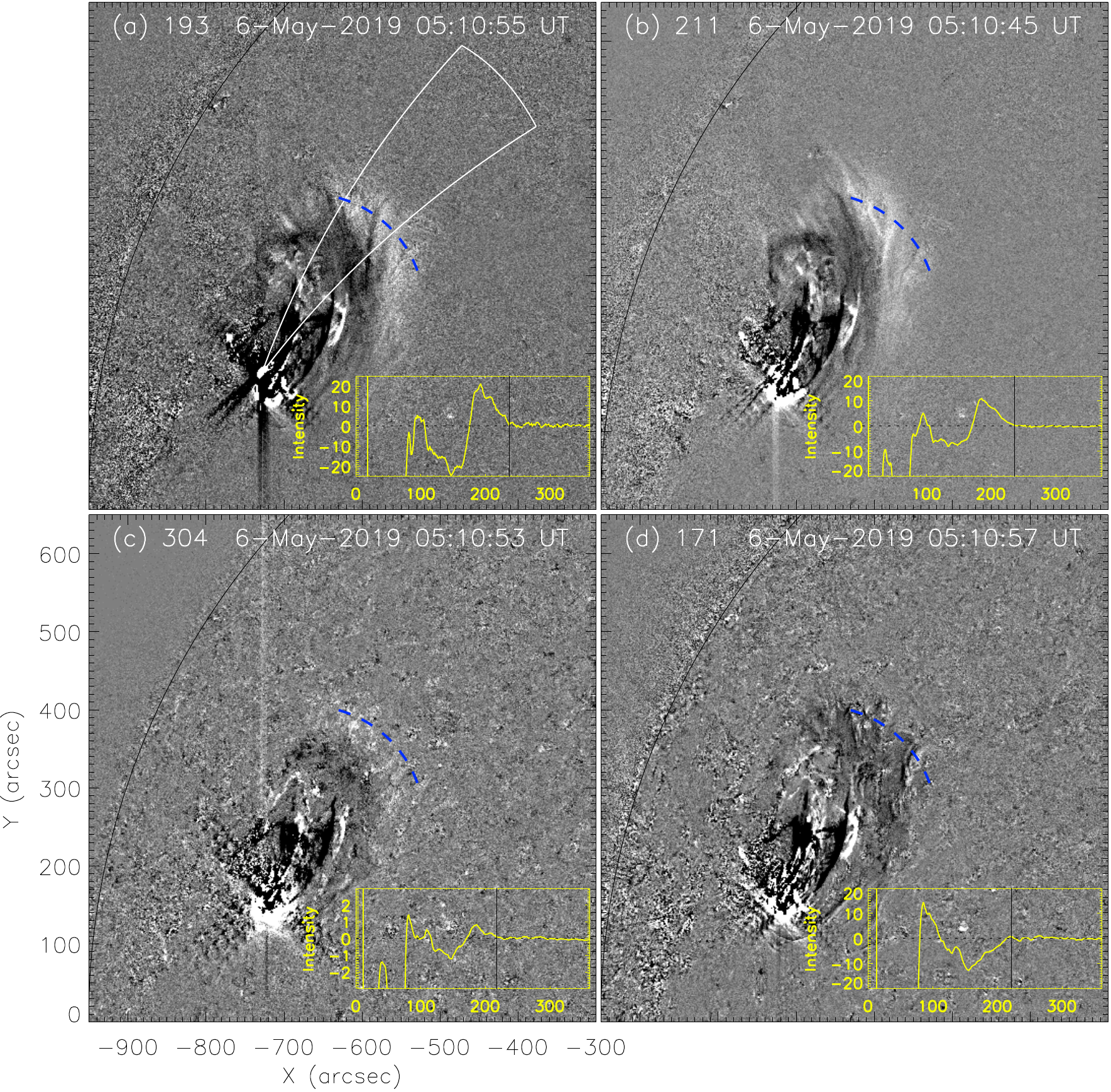}
\caption{The signatures of EUV wave in different passband at around 05:11 UT. (a) 193 $\rm\AA$ running difference image. (b) 211 $\rm\AA$ running difference image. (c) 304 $\rm\AA$ running difference image. (d) 171 $\rm\AA$ running difference image. The blue dashed arc-shape dashed lines indicate the corresponding Moreton wavefront identified in the panel (d) of Fig.\ref{fig5}. The white fan-shape in panel (a) denotes the same sector as Moreton wave. The yellow curves in the bottom right of different panels show the variations of intensity of different passbands along the sector. The horizontal dotted lines denote the zero level of different intensity. The black curves in different panels denote the solar limb.}\label{fig7}
\end{figure}

Fig.\ref{fig8} shows the evolution of the EUV wave in running-difference images of 193 $\rm\AA$. The EUV wave displayed a similar characteristic behaviour with the Moreton wave, which originated in the small-scale filament eruption site and propagated toward the solar north. Many fascinating features could be found. At 05:12:40 UT, some parts of the wave underwent refracting at the site marked by the yellow arrow in the panel (b). When the wave went through the place nearby the site marked by the yellow arrow, it became slower and changed its direction of propagation toward the southwest of the Sun (see the animation of Fig.\ref{fig8}). Comparing the wavefronts before refracted and after refracted marked in panel (h), the direction of this part of the wave had changed a lot}、. The yellow arrows in panels (b)-(h) mark the evolution of this part of the wave and we define it as refracted wave. At the other place, some parts of the wave underwent diffracting, which was marked by the blue arrows in the panel (b)-(f). These parts of the wave gradually changed it wavefront direction marked by the blue dotted line in panels (b)-(h) and we define it as diffracted wave. It is reasonable to suspect that a specific coronal medium would exist nearby the site marked by the yellow arrow in the panel (b), which caused the wave diffracting. By using PFSS extrapolation, we found that there was a magnetic separtrix in there showing in the panel (c) of Fig.\ref{fig11}. At 05:13:52 UT, a wave propagating toward the northeast direction appeared nearby the refracted wave, marked by the red arrow in the panel (c). Red arrows in panels (c)-(g) marked this wave propagation and we name it as reflected wave. The blue, yellow, and red arrows indicate the evolution of the diffracted, refracted and reflected waves, respectively. These waves generally fainted and disappeared after 05:23:04 UT. More details in the evolution of the EUV wave would be displayed in the animation of the Fig.\ref{fig8}.

\begin{figure}[t!]
\figurenum{8}
\plotone{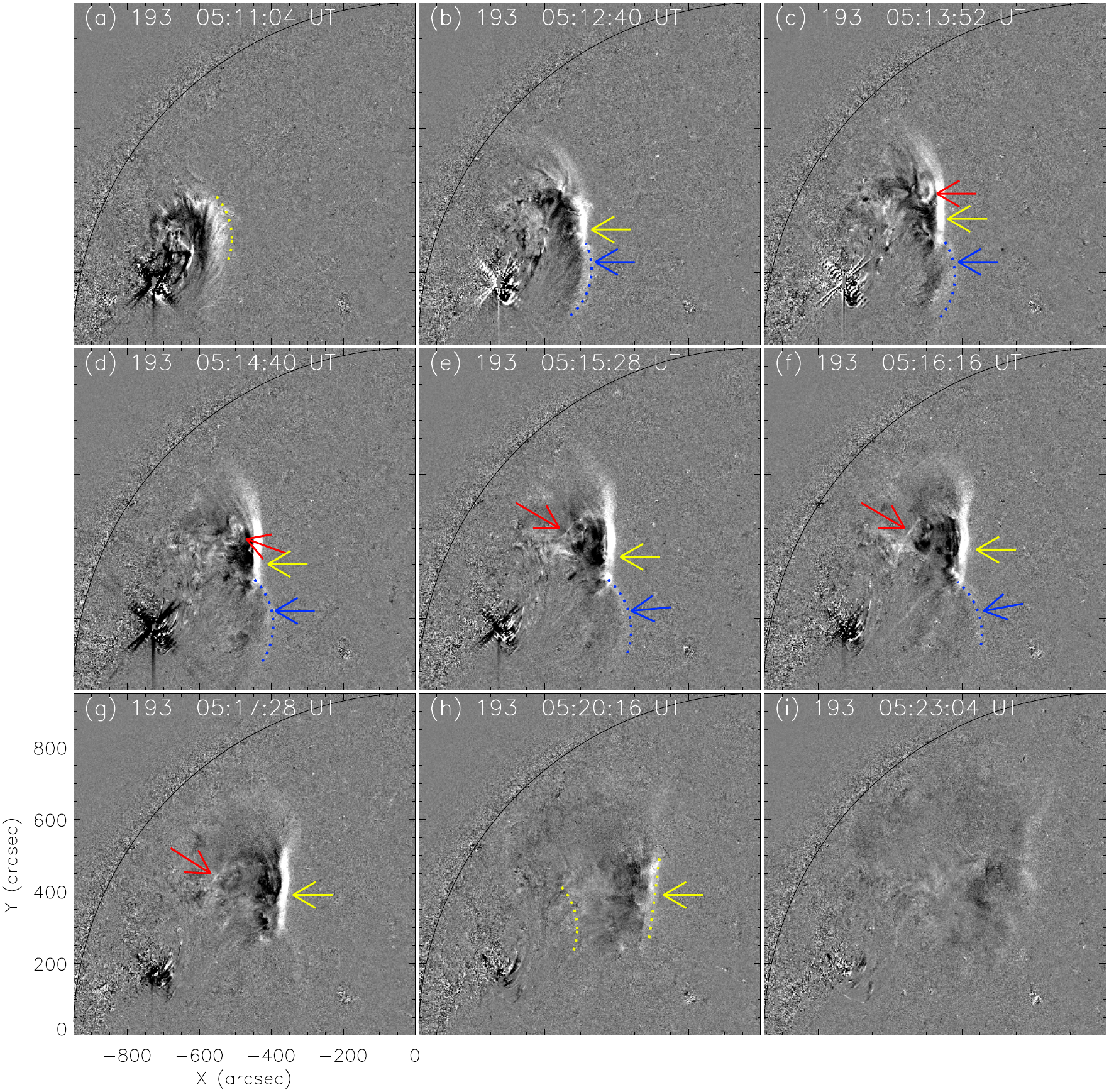}
\caption{Evolution of the EUV wave in 193 $\rm\AA$ running difference images. The yellow arrows indicate the evolution of the refracted wave while the blue arrows mark the aspect of the diffracted wave. The reflected wave can be identified by the red arrows. The black curves in different panels depict the solar limb. The yellow dashed line in panel (a) depicts the wavefront before the wave refracted. The right yellow dashed line in panel (h) depicts the wavefront after the wave refracted and the other yellow dashed line is the same as the one in panel (a). The blue dashed lines mark the diffracted wave in panels (b)-(f).}\label{fig8}
\end{figure}

\begin{figure}[t!]
\figurenum{9}
\plotone{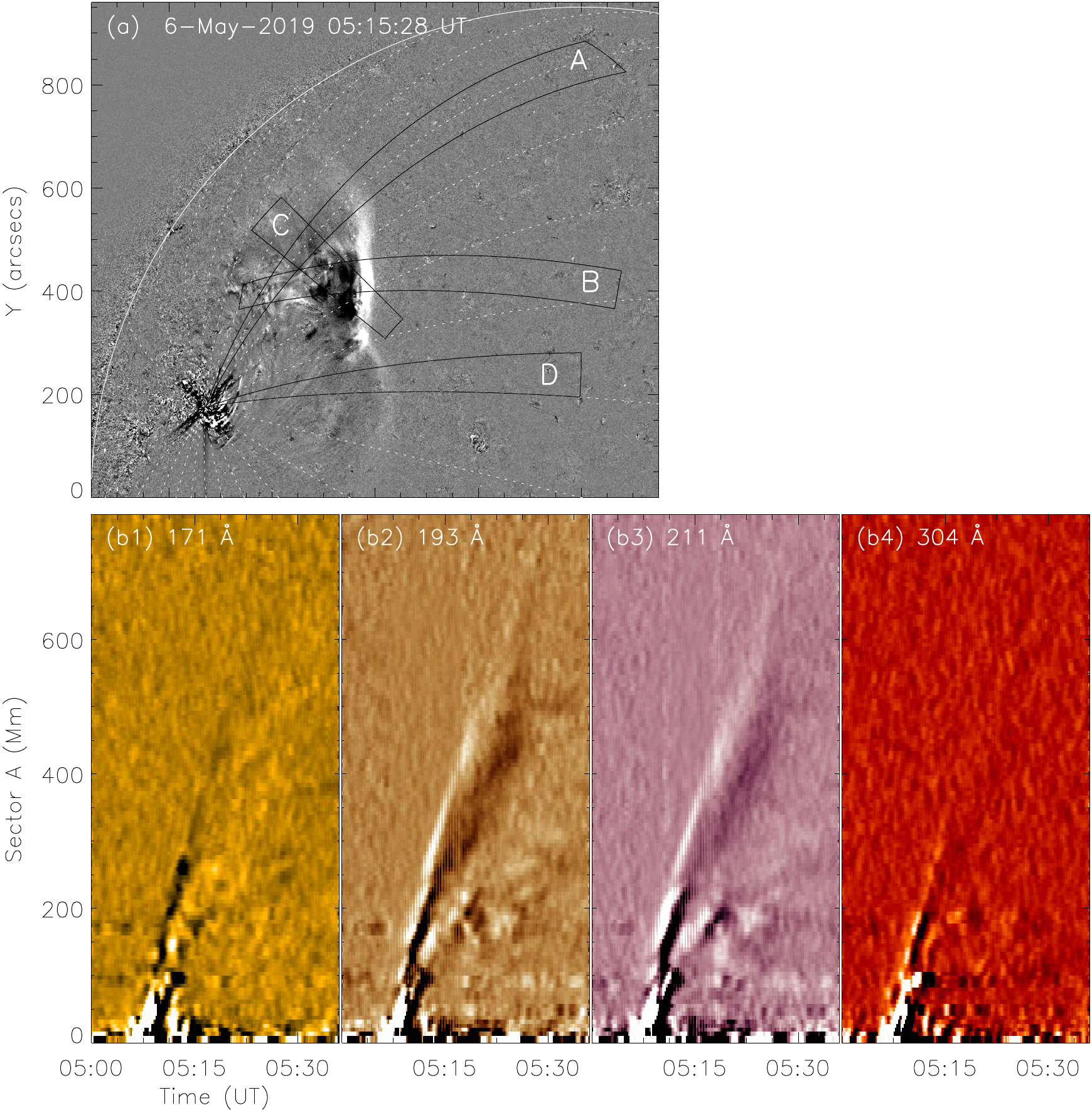}
\caption{(a) SDO/AAI 193 $\rm\AA$ images at 05:14:16 UT. The white dotted lines is the ``corresponding longitude" of the new solar coordinates system. Positions of the Sector A, Sector B, Sector C and Sector D are marked by the black arc-shape lines. The white solid line denotes the solar limb. (a1)-(a4) The stack plots derived by the running difference images in 171 $\rm\AA$, 193 $\rm\AA$, 211 $\rm\AA$ and 304 $\rm\AA$. The blue dashed lines in different panels plot the slope of EUV signatures in different passbands.}\label{fig9}
\end{figure}

When the wave propagated freely in the quiet Sun and did not effected by any specific coronal medium and magnetic structure, it would show the natural characteristics of the EUV wave's propagation and we name it as unaffected wave. In order to determine the kinematic characteristics of these waves (including unaffected, refracted, reflected and diffracted waves), we make four sectors: Sector A, Sector B, Sector C and Sector D. Sector A and Sector C are set as the similar method as the sector in the panel (a) of Fig.\ref{fig6}, which the new “north pole” is also set at the eruption center (x=-728$\arcsec$, y=174$\arcsec$). The angular extent of Sector A is 10$\degr$ while the one of Sector C is 7 $\degr$. On the other hand, the Sector B and Sector D are set to make it perpendicular to the wavefronts of refracted and reflected waves, respectively. It should keep in mind that coronal shock wavefronts are observed as a line-of-sight integration of optically thin emission and projection effects are involved in the signatures of the wavefronts. In addition, the shock wave observed by different passbands only represents the wave disturbance in different temperature plasma response. Therefore, we determine the dynamic characteristics of the EUV wave \textbf{by using the stack technique} that the measurement are corrected for spherical solar surface as previous studies \citep{liu10}, which assumes that all emission originates from the spherical solar surface and would \textbf{overestimate the measurements of position and speed of the signature at a constant height according to the location of the wave} \citep{liu14w}. The detail analysis of error estimation is given at Appendix.\ref{appendix}. Each data point in stack plots is the average of the pixels within a sector along a circular arc at the same ``corresponding latitudinal". Thus, Sectors A, B, C, and D are utilized to analyze the kinematic properties of unaffected, refracted, reflected and diffracted waves, respectively. Panels (b1)-(b4) of Fig.\ref{fig9} exhibit stack plots of running difference images in 171 $\rm\AA$, 193 $\rm\AA$, 211 $\rm\AA$ and 304 $\rm\AA$ along the Sector A, respectively. The unaffected wave shown a similar tendency of the propagation in different EUV passbands. The signature of this wavefront in 193 $\rm\AA$ and 211 $\rm\AA$ is more distinct than the one in the other two passbands. Based on the panels (b2) \& (b3), this wave could propagate away up to about 700 Mm far away from the eruption center.

\begin{figure}[t!]
\figurenum{10}
\plotone{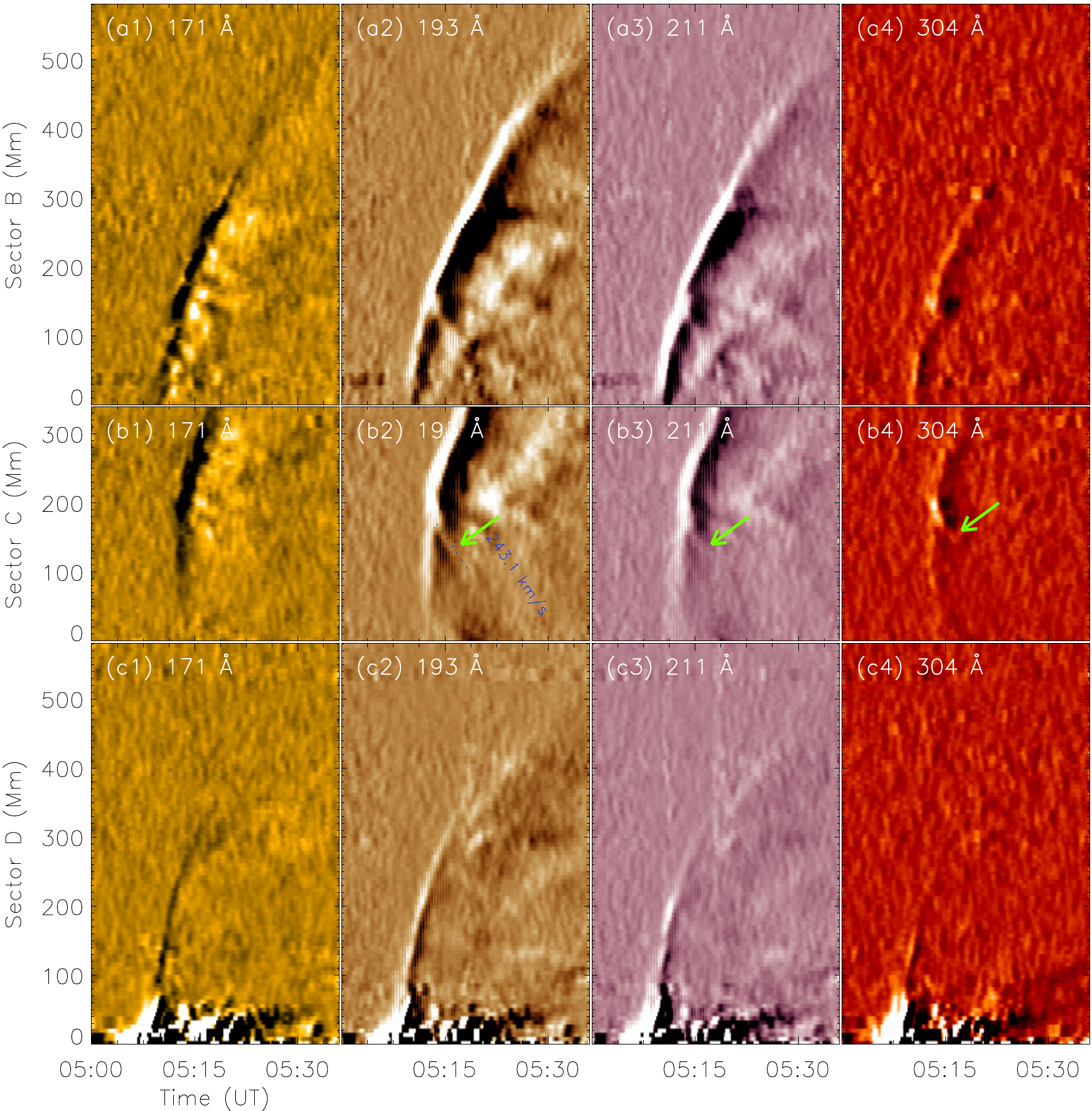}
\caption{The stack plots tracing the refracted and diffracted waves a along the Sectors B, C and D. (a1)-(a4) The stack plots along the Sector B derived by the running difference images of 171 $\rm\AA$, 193 $\rm\AA$, 211 $\rm\AA$ and 304 $\rm\AA$, respectively. (b1)-(b4) The stack plots along the Sector C derived by the runnning-difference images of 171 $\rm\AA$, 193 $\rm\AA$, 211 $\rm\AA$ and 304 $\rm\AA$, respectively. \textbf{The signature of reflected wave is indicated by green} arrows in panels (b2)-(b4). The blue dashed line and blue text are the identified trace and the velocity of reflected wave, respectively. (c1)-(c4) The stack plots along the Sector D derived by the runnning difference images of 171 $\rm\AA$, 193 $\rm\AA$, 211 $\rm\AA$ and 304 $\rm\AA$, respectively.}\label{fig10}
\end{figure}

The results of the stack plots along Sectors B, C and D have been depicted in Fig.\ref{fig10}. Figs.\ref{fig10} (a1)-(a4) are the stack plots of running difference images in 171 $\rm\AA$, 193 $\rm\AA$, 211 $\rm\AA$ and 304 $\rm\AA$ along the Sector B, while panels (b1)-(b4) and (c1)-(c4) are the stack plots along Sectors C and D, respectively. The wavefronts of refracted, reflected, and diffracted waves also shown corresponding signatures in four EUV passbands, which traced along the Sectors B, C and D. Alike the stack plots of Sector A, the signatures in 193 $\rm\AA$ and 211 $\rm\AA$ are more distinct than the one in the other two passbands. As is the exhibition of the stack plots (a1)-(a4) and (c1)-(c4), the refracted and diffracted waves showed a decelerating profiles. This means the refracted and diffracted waves underwent a decelerating propagation along Sectors B and D. On the other hand, We can also distinctly identify the signatures of reflected wave in the 193 $\rm\AA$, 211 $\rm\AA$ and 304 $\rm\AA$ stack plots, which is marked by the green arrows in the panels (b2) - (b4). The velocity of the reflected wave can be derived by the signature in panel (b2), which is estimated to be about 241.1 $\pm$ 30.8 km s$^{-1}$.

\begin{figure}[ht!].
\figurenum{11}
\plotone{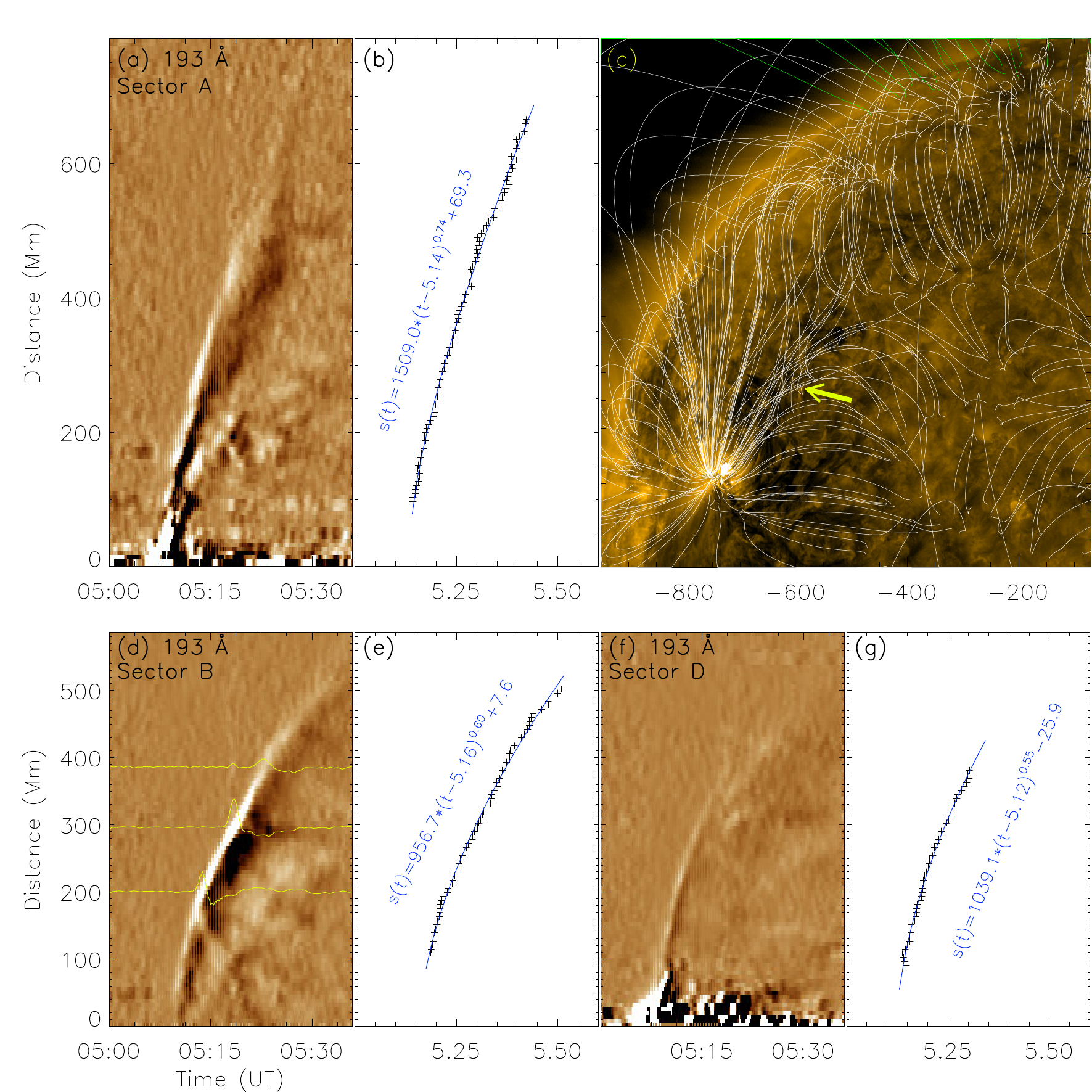}
\caption{(a) The stack plot along Sector A by using 193 $\rm\AA$ running difference image. (b) The identified position of the unaffected wave. The blue line is the power-law fitting profile. (c) SDO/AIA 171 $\rm\AA$ image superimposed by the extrapolated magnetic fields derived by the Potential Field Solar Surface model. The white lines denote the magnetic field lines. The background is SDO/AIA 171 $\rm\AA$ image. The yellow arrow marks the position of a magnetic separatrix. (d) The stack plot along Sector B by using 193 $\rm\AA$ running difference image. The yellow dotted line indicates the path of reflected wave. The yellow solid lines denote the intensities profiles at different sites as a function of time. (e) The identified position of the refracted wave. The blue line is the power-law fitting. (f) The stack plot along Sector C by using 193 $\rm\AA$ running difference image. (g) The identified position of the diffracted wave. The blue line is the line of power-law fitting.}\label{fig11}
\end{figure}

In order to investigate the kinematic characteristics of the unaffected, refracted, and diffracted waves in detail, we meticulously identify the wavefronts along the Sectors A, B and D. We use the stack plots derived by the running difference image in 193 $\rm\AA$ for analyzing due to vivid signatures of different waves of 193 $\rm\AA$ stack plots (see Fig.\ref{fig11} (b), (e) and (g)). We determine the wave position by using the average position of the beginning enhancement and the maximum of wave signature. As is shown in panel (d), every row in the stack plot could determine a moment position of the wave by using above method. Therefore, the whole propagating path of the wave can be derived by the whole stack plot. The tracked propagating paths of unaffected, refracted and diffracted waves are plotted in panels (b), (e) and (g), respectively. It is found that these waves propagated with deceleration with a non-constant rate. Therefore, we use power-law function to fit their propagating trajectories. The power-law is given by:
\begin{equation}\label{eq1}
   s(t)=c_1(t-t_0)^{\kappa}+c_2,
\end{equation}
where $t_0$ is given an arbitrary value associated with the initial time. As is well known, their speed and acceleration function are evaluate by the first and second order differential function, respectively, which are following equations:
\begin{equation}\label{eq2}
   v(t)=c_1 \kappa (t-t_0)^{\kappa-1},
\end{equation}

\begin{equation}\label{eq3}
   a(t)=c_1 \kappa (\kappa-1) (t-t_0)^{\kappa-2}.
\end{equation}

Table \ref{tab1} presents the results of the fitting described above. The errors of parameters are the fitting errors. On the other hand, according to the analysis in the Appendix, \textbf{we also corrected the fitting function when considering the error of the overestimation in the measurement}. Unaffected wave exhibited a smaller deceleration than the other two waves, 、which is close to propagation with a constant speed. According to the acceleration functions, the refracted and diffracted waves underwent a decelerated propagation with a relative bigger deceleration rate. Using these corrected analytic functions with time, the speed and deceleration of refracted or diffracted waves can be derived at any moment. Table \ref{tab2} shows the kinematic results of different waves according to these corrected analytic functions. Such as, at 5:15 UT ($t=5.25$), the speed and deceleration rate of unaffected wave were 538.3 $\pm$ 21.6 km s$^{-1}$ and -0.35 $\pm$ 0.03 km s$^{-2}$, while the ones of refracted wave were 401.3 $\pm$ 18.3 km s$^{-1}$ and -0.50 $\pm$ 0.04 km s$^{-2}$ and the ones of diffracted wave were 381.2 $\pm$ 34.2 km s$^{-1}$ and -0.37 $\pm$ 0.05 km s$^{-2}$. At 5:18 UT ($t=5.30$), the speed and deceleration rate of unaffected wave were 488.3 $\pm$ 17.8 km s$^{-1}$ and -0.22 $\pm$ 0.02 km s$^{-2}$, while the ones of refracted wave became 336.3 $\pm$ 13.9 km s$^{-1}$ and -0.27 $\pm$ 0.02 km s$^{-2}$ and the ones of diffracted wave became 329.1 $\pm$ 27.4 km s$^{-1}$ and -0.23 $\pm$ 0.03 km s$^{-2}$. Moreover, according the derived functions of unaffected wave, the propagation speed of the EUV wave was about 686.1 $\pm$ 34.0 km s$^{-1}$ at around 05:11 UT ($t=5.18$). At around this time, we have obtained that the propagation speed of Moreton wave was about 611.8 $\pm$ 41.2 km s$^{-1}$ (see the analysis of Moreton wave in this section). Comparing with the propagation speeds of EUV wave and Moreton at that moment, it can derived that unaffected EUV wave propagated faster than Moreton wave. Therefore, it is reasonable that the EUV wavefronts were in front of the Moreton wavefront at around 05:11 UT (see Fig.\ref{fig7}). Additionally, we also obtained the propagation speed of EUV wave was about 591.0 $\pm$ 25.9 km s$^{-1}$ at around 05:13 UT when the reflected wave begin to appear.

\begin{deluxetable*}{cCCC}
\tablecaption{Dynamic characteristics of different waves by using power-law fitting \label{tab1}}
\tablecolumns{4}
\tablenum{1}
\tablehead{
\colhead{ } &
\colhead{unaffected wave} &
\colhead{refracted wave} &
\colhead{diffracted wave}
}
\startdata
$s(t)$ & (1509.0$\pm$7.1)*(t-5.14)^{0.74$\pm$0.01} & (956.7$\pm$4.7)*(t-5.16)^{0.60$\pm$0.01} &(1039.1$\pm$13.2)*(t-5.12)^{0.55$\pm$0.02}  \\
&+69.3$\pm$1.5 &+7.6$\pm$4.6&-25.9$\pm$7.2\\
corrected $s(t)$ & (1474.3$\pm$6.9)*(t-5.14)^{0.74$\pm$0.01} & (918.4$\pm$4.5)*(t-5.16)^{0.60$\pm$0.01} & (993.4$\pm$12.6)*(t-5.12)^{0.55$\pm$0.02} \\
&+67.7$\pm$1.5&+7.3$\pm$4.4&-24.8$\pm$6.9 \\
\hline
corrected $v(t)$ & (1091.0$\pm$19.8)*(t-5.14)^{-0.26$\pm$0.01} & (551.0$\pm$11.9)*(t-5.16)^{-0.40$\pm$0.01} & (546.4$\pm$26.8)*(t-5.12)^{-0.45$\pm$0.02}\\
corrected $a(t)$ & (-283.7$\pm$16.1)*(t-5.14)^{-1.26$\pm$0.01} &(-220.4$\pm$10.8)*(t-5.16)^{-1.40$\pm$0.01} &(-245.9$\pm$23.0)*(t-5.12)^{-1.45$\pm$0.02}
\enddata
\tablecomments{The unit of time is hour (h) while the one of distance is megameter (Mm)}
\end{deluxetable*}

\begin{deluxetable*}{cccccccccccccccc}
\tablenum{2}
\tablecaption{The kinematic results of different waves base on corrected functions. \label{tab2}}
\tablehead{
\colhead{Time} & & \multicolumn{2}{c}{unaffected wave}& & \multicolumn{2}{c}{refracted wave} & & \multicolumn{2}{c}{diffracted wave} \\
\colhead{(UT)} & & \colhead{speed} & \colhead{acceleration}& & \colhead{speed} & \colhead{acceleration} & & \colhead{speed} & \colhead{acceleration}
}
\startdata
05:11 & & 686.1 $\pm$ 34.0 & -1.15 $\pm$ 0.10 & & * & * & & * & * \\
05:13 & & 591.0 $\pm$ 25.9 & -0.56 $\pm$ 0.04 & & 483.0 $\pm$ 24.3 & -0.95 $\pm$ 0.07 & & 435.8 $\pm$ 41.6& -0.56 $\pm$ 0.08 \\
05:15 & & 538.3 $\pm$ 21.6 & -0.35 $\pm$ 0.03 & & 401.3 $\pm$ 18.3 & -0.50 $\pm$ 0.04 & & 381.2 $\pm$ 34.2 & -0.37 $\pm$ 0.05 \\
05:18 & & 488.3 $\pm$ 17.8 & -0.22 $\pm$ 0.02 & & 336.3 $\pm$ 13.9 & -0.27 $\pm$ 0.02 & & 329.1 $\pm$ 27.4 & -0.23 $\pm$ 0.03 \\
05:25 & & 423.4 $\pm$ 13.1 & -0.11 $\pm$ 0.01 & & 263.8 $\pm$ 9.3 & -0.11 $\pm$ 0.01 & & *&*
\enddata
\tablecomments{The unit of speed is km s$^{-1}$ while the one of acceleration is km s$^{-2}$.}
\end{deluxetable*}

Fig.\ref{fig11} (c) exhibits the magnetic topological structures derived from the potential field source surface (PFSS) model \citep{alt69,sch69}. We can find an intriguing feature that an interface between two different magnetic systems (or magnetic separatrix) inhabit in the place marked by the yellow arrow in the panel \textbf{(c)} of Fig.\ref{fig11}. We suspect this magnetic separatrix is a key factor for the refracting, reflecting and diffracting of the EUV wave. When the EUV shock wave interacted with this magnetic separatrix, it would cause this EUV shock wave to refract, reflect and diffract. Influenced by the magnetic separatrix, some parts of wave went through the magnetic separatrix with a larger deceleration and changed its direction of propagation, while some parts of wave would be reflected by the magnetic separatrix. When the ambient part of wave interacted with this magnetic separatrix, it would cause this part of wave to diffract. Therefore, the inhomogeneities of the coronal magnetic structure would effect the propagation of the EUV wave.

\section{Summary and discussion}\label{sec:conclusion}
In this paper, we study a small-scale filament by using high spatial resolution observations from NVST and analyze some global wave-like perturbations (Moreton and EUV waves) induced by this small-scale filament eruption. The main results are as follows.

(1) A small-scale filament gradually appeared under the filamentary arcade. Some elongate dark lanes or filamentary structures on the photosphere became lighter and moved toward south. Both of signatures demonstrated that the small-scale filament experienced an emerging and lifting motion.

(2) The small-scale filament eruption triggered a narrow CME with angle width 52 degrees and mean propagating speed of 376 km s$^{-1}$. After the bright CME front ejected out, the shrinking motion of the magnetic field was found.

(3) The small-scale filament eruption simultaneously induced the Moreton and EUV waves. The Moreton wave propagated toward solar north with the speed of $\sim$ 611.8  $\pm$ 41.2 km s$^{-1}$. The similar morphologies of Moreton and EUV wave fronts in different passbands were derived. Therefore, we conclude that Moreton wave and EUV waves in different passbands are perturbations in different layer triggered by the same shock wave.

(4) The EUV wave propagated with a non-constant deceleration rate along an non-effected path in the corona and the speed function was $(1091.0\pm19.8)*(t-5.14)^{-0.26\pm0.01}$. At around 05:11 UT, the unaffected EUV wave propagated with the speed of about 686.1 $\pm$ 34.0 km s$^{-1}$, which was faster than Moreton wave of 611.8 $\pm$ 41.2 km s$^{-2}$.

(5) The reflected, refracted and diffracted waves were found during the propagation of the EUV wave. We obtain the refracted wave propagating function of $(918.4\pm4.5)*(t-5.16)^{0.60\pm0.01}$ and the diffracted wave propagating function of $(993.4\pm12.6)*(t-5.12)^{0.55\pm0.02}$. Using the PFSS model extrapolation, a magnetic separatrix was found nearby the position of reflected, refracted and diffracted waves. Therefore, we consider the magnetic separatrix should be responsible for causing the EUV wave reflecting, refracting and diffracting.

The propagating direction of Moreton and EUV waves was toward the solar north. This may be related to the morphology of the small-scale filament eruption and the existence of the large sunspot in the western of the small-scale filament. The like blowout-type eruption of the small-scale filament resulted in the ejected material along the magnetic field lines from the large sunspot ambient to solar north. The magnetic field strength also is decreasing from the large sunspot ambient to solar north. Moreover, shock waves are always launched into the direction away from the active region (e.g., \citealp{war04,xue13,war15}). The topology of these magnetic fields would set favorable conditions for the propagation of a coronal magneto-hydrodynamic shock wave along the magnetic field direction. Furthermore, the strong magnetic pressure of the sunspot probably restraint the waves propagating toward southwestern \citep{uch73,zha11}. Therefore, under these situations, both Moreton and EUV waves propagated in a restricted direction (toward the solar north) instead of the symmetrical circle expansion.

The joint observations of Moreton and EUV waves are relatively rare. Since the SDO launched, only a few Moreton and EUV waves could be simultaneous captured and studied. \cite{asa12} presented the simultaneous observation of an H$\alpha$ Moreton wave, a corresponding EUV fast coronal waves, and a slow and bright EUV wave. They derived the almost equivalent propagation speeds of the Moreton wave and the fast EUV wave along the same path (760 km s$^{-1}$ and 730 km s$^{-1}$). Using by very high cadence H$\alpha$ observations from the Halpha Solar Telescope for Argentina (HASTA) and SDO observations, \cite{fra16} analyzed the correlation between a Moreton wave and the EUV coronal bright fronts occurring on 29 March 2014, and found that the Moreton and 304 $\rm\AA$ EUV waves displayed a similar propagation with a non-constant deceleration. Using the data of the Flare Monitoring Telescope (FMT), \cite{cab19} studied the same event as \cite{fra16} and found that the downward motion of the chromospheric material at the Moreton wave could be up to 4 km s$^{-1}$. \cite{lon19} studied four homologous global waves (including EUV and Moreton waves) originating from the same active region, and found the EUV waves inhibited high initial velocity and strong deceleration which exceeded those of the Moreton waves. In our study, we also present the simultaneous observations of the Moreton and EUV waves induced by a small-scale filament eruption. We obtain the propagation speed of Moreton wave is about 611.8 $\pm$ 41.2 km s$^{-1}$ and that of EUV wave is about 686.1 $\pm$ 34.0 km s$^{-1}$ at the corresponding time. It is reasonable that wavefronts in the EUV passbands propagated further distances than the Moreton wave until around 05:11 UT (see Fig.\ref{fig7}). This is consistent with the results found by \cite{vrs16}, where the chromospheric perturbation lags behind the transition-region and coronal perturbation in their numerical simulation of coronal large-amplitude wave. Therefore, we consider that different passbands EUV and Moreton waves are the perturbation signals in different layer induced by the same dome shock.

On the other hand, the EUV wave underwent reflecting, refracting and diffracting when it went through a magnetic separatrix. These features are strong indications for the true wave nature of the disturbances. And these features are also consistent with numerical simulations treating EUV wave as fast-mode shock waves, which exhibits that the EUV wave would experience strong refraction and deflection when it interacts with active regions and coronal holes \citep{wan00,wu01,ofm02}. Furthermore, we derive the propagation speed of the EUV wave is about 400-700 km s$^{-1}$, which is consistent with the propagating speed of a fast-mode shock wave in the corona \citep{gal11,war15}. Therefore, we consider that the EUV wave in our study would be a signal of a fast-mode shock wave instead of a pseudo wave by successive stretching magnetic field lines (e.g., \citealp{del99,che02}).

Previous studies also found that coronal bright fronts propagated with a deceleration (e.g., \citealp{war05,ver08,xue13,fra16}). In our study, we have also found that the unaffected coronal shock wave exhibited a deceleration propagation with a decreasing rate. This is consistent with a nonlinear fast-mode shock wave, where the propagation speed of the shock wave in a homogeneous and isotropic environment theoretically depends on the Mach number or wave’s amplitude (e.g., \citealp{pri14,vrs16}). As the shock wave propagates, its amplitude decreases due to the geometric expansion of its front and the fact that its leading edge is propagating faster than the trailing edge.
The amplitude decrease results in a deceleration of the shock wave. Additionally, a power-law fit is a better representation of the distance-time curves of freely propagating shock waves \citep{war11} according the the Sedov solution \citep{sed59}. On the other hand, we found the speed of the reflected wave (243.1 $\pm$ 30.8 km s$^{-1}$) is much smaller than the speed of the incident wave (which approximately equals to the one of unaffected wave at around 05:13 UT: 591.0 $\pm$ 25.9 km s$^{-1}$). In other words, the reflected wave was slower than the primary wave. There may be several possibilities for this feature. The primary wave is a nonlinear wave or shock, which is propagating with a Mach number larger than unity, while the reflected wave had changed to be a linear one traveling at the characteristic speed \citep{liu14w,war15}. The reflected wave has to propagate in the rest frame of flow field associated with the primary wave, which may result in an apparently lower speed \citep{kie13}. Additionally, it may relate to the properties of medium (magnetic separatrix) causing the shock wave to reflect. The compressible magnetic separatrix may cause the decreasing of the reflected wave speed, when the shock wave reacted with the magnetic sparatrix. Moreover, it is interesting to find that the refracted and diffracted wave underwent a deceleration with bigger non-constant deceleration rate. We obtain that their deceleration rate was decreasing with time, which likes a damping motion. These phenomena would help us to further understand the nature of the interaction between magneto-hydrodynamic shock wave and local magnetic field.

\acknowledgments
The authors thank the referee for constructive suggestions and comments that helped to improve this paper. SDO is a mission of NASA's Living With a Star Program. The authors are indebted to the SDO, GONG, NVST teams for providing the data. We thank Wei Liu for helpful suggestions and discussions. This work is supported by the National Science Foundation of China (NSFC) under grant numbers 11873087, 11803085, 11603071, 11633008, the Yunnan Talent Science Foundation of China (2018FA001), the Yunnan Science Foundation of China (2019FD085), the CAS ``Light of West China" Program under number Y9XB018001, the Open Research Program of the Key Laboratory of Solar Activity of Chinese Academy of Sciences (grant No.
KLSA202014), the grant associated with project of the Group for Innovation of Yunnan province.

\appendix

\section{error estimations}\label{appendix}
In this study, we determine the dynamic characteristics of EUV wave by using the method as \cite{liu10}. This track techniques neglect heights of EUV waves by assuming that all emission originates from the spherical solar surface and this measurement of distance was corrected by the sphericity of the solar photosphere. This method would cause the overestimation in measurement of some parameters(e.g., distance, velocity). We will supply some error estimations in the coming sections.

\begin{figure}[t]
\plotone{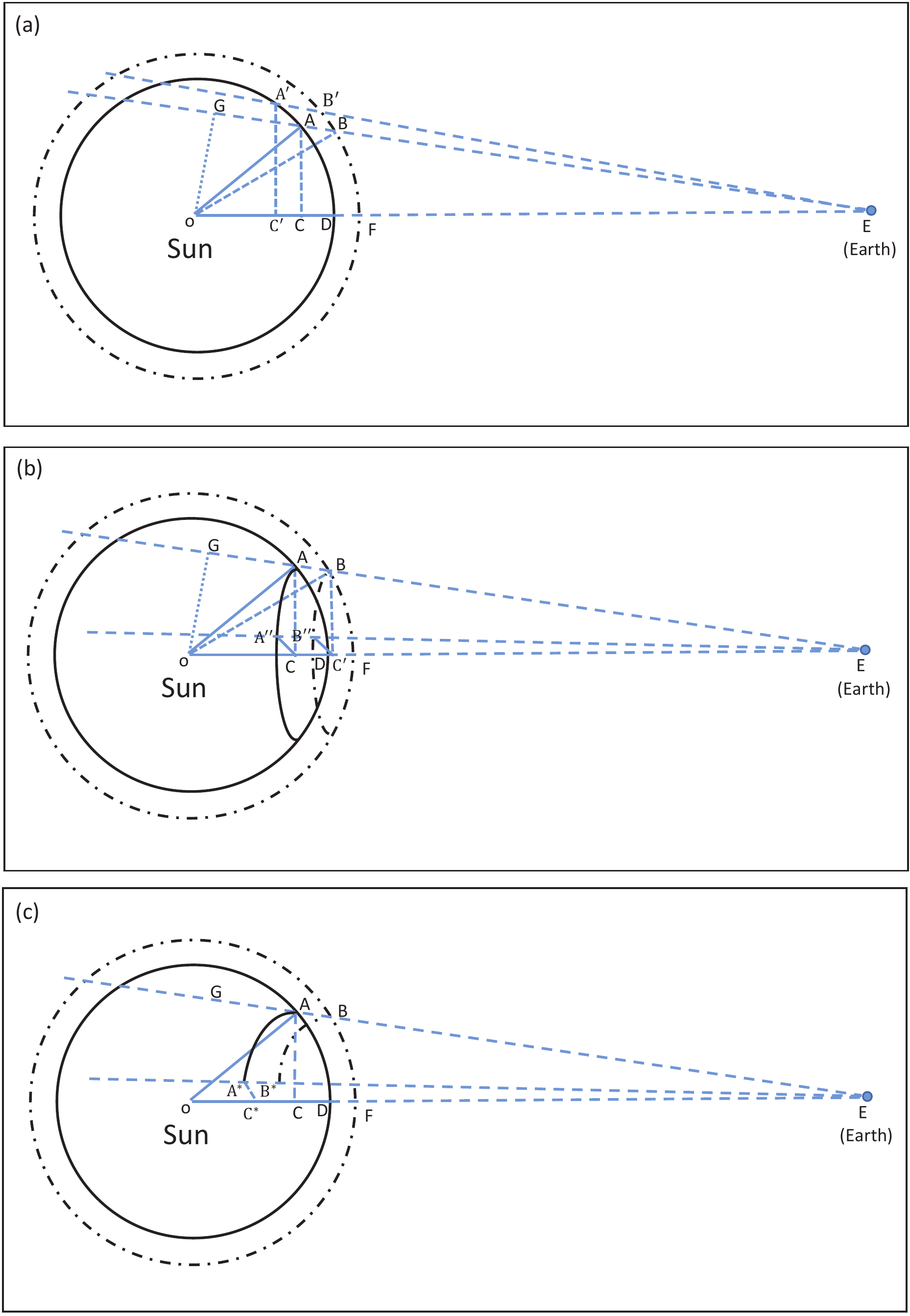}
\caption{Solid and dotted-dashed circles denote the Sun surface and constant height corona, respectively.}\label{fig12}
\end{figure}

As Fig.\ref{fig12} (a) shown, a coronal structure B in the height of $\Delta h$ above the solar surface, projects onto the solar surface A observed by a observer on the Earth (E). We set new “north pole” at the disk center (D) and \textbf{make AC = $d$,} AC $\perp$ OE, OA = OD = R$_\sun$, OB = OF = R$_\sun+\Delta h$, DE =1 AU. \textbf{Defining $\theta$ = $\angle \rm AEO$, we can get}
 \begin{eqnarray}
 \angle \rm AOC &=& \arcsin(\frac{\rm AC}{\rm OA}) \nonumber\\
                &=& \arcsin(\frac{d}{\rm R_\sun}) \label{a1},\\
         \rm OC &=& \sqrt{\rm OA^2 - \rm AC^2} \nonumber\\
                &=& \sqrt{\rm R_\sun^2 - \it d \rm^2}\\
         \theta &=& \arctan(\frac{\rm AC}{\rm CE}) \nonumber\\
                &=& \arctan(\frac{\rm AC}{\rm CD+\rm DE}) \nonumber\\
                &=& \arctan(\frac{\rm AC}{\rm OD-OC+DE}) \nonumber\\
                &=& \arctan(\frac{d}{\rm R_\sun-\sqrt{R_\sun^2 - \it d \rm^2} + 1 AU}) \label{a3}.
\end{eqnarray}

We make OG $\perp$ EG and get
 \begin{eqnarray}
 \rm OG &=& \rm OE * \sin(\angle \rm AEO) \nonumber \\
        &=& \rm (OD+DE)* \sin(\angle \rm AEO)\nonumber \\
        &=& \rm(R_\sun + 1 AU)*\sin(\theta)\\
 \rm GE &=& \rm OE * \cos(\angle \rm AEO)\nonumber \\
        &=& \rm(R_\sun + 1 AU)*\cos(\theta)\\
 \rm GB &=& \rm \sqrt{OB^2-OG^2}\nonumber \\
        &=& \rm \sqrt{(R_\sun +\Delta \it h \rm)^2-[(R_\sun + 1 AU)*\sin(\theta)]^2}.
 \end{eqnarray}
Then, we get
 \begin{eqnarray}
 \rm BE &=& \rm GE -GB \nonumber\\
        &=& \rm(R_\sun + 1 AU)*\cos(\theta) - \rm \sqrt{(R_\sun +\Delta \it h \rm)^2-[(R_\sun + 1 AU)*\sin(\theta)]^2}.\label{a7}
 \end{eqnarray}
According to the law of sines, in the $\triangle$ BOE, we can get
\begin{eqnarray}
 \rm \frac{OB}{\sin \angle AEO} &=&\rm \frac{BE}{\sin \angle BOE} \\
 &\Downarrow& \nonumber \\
 \rm \angle BOE &=& \rm \arcsin(\frac{BE}{OB}*\sin \angle AEO) \label{a9}.
\end{eqnarray}
Based on above equations, we can get
\begin{eqnarray}
 \rm \stackrel{\frown}{AD} &=& \rm OA*\angle AOC \nonumber\\
        &=& \rm R_\sun * \arcsin(\frac{\it d}{\rm R_\sun}) \label{a10}\\
 \rm \stackrel{\frown}{BF} &=& \rm OB*\angle BOE \nonumber \\
        &=& \rm (R_\sun + \Delta h)*\beta \label{a11},
\end{eqnarray}
where
\begin{eqnarray}
  \beta&=&\arcsin\{\frac{\rm(R_\sun + 1 AU)*\cos(\theta) - \rm \sqrt{(R_\sun +\Delta \it h \rm)^2-[(R_\sun + 1 AU)*\sin(\theta)]^2}}{(R_\sun + \Delta h)}*\sin(\theta)\}\nonumber \\
  \theta &=& \arctan(\frac{d}{\rm R_\sun-\sqrt{R_\sun^2 - \it d \rm^2} + 1 AU}).\nonumber
\end{eqnarray}

Firstly, if the EUV wave originally propagates along the ``corresponding longitude" from B to $\rm B\arcmin$ taking time of t, which projects onto the solar surface from A to A$\arcmin$ (see Fig.\ref{fig12} (a)), then the different distance ($\Delta l\arcmin$) could be expressed as
\begin{eqnarray}
  \Delta l\arcmin&=& |\rm \stackrel{\frown}{A\arcmin A}-\stackrel{\frown}{B\arcmin B} | \nonumber \\
          &=& |\rm (\stackrel{\frown}{A\arcmin D}-\stackrel{\frown}{AD}) - (\stackrel{\frown}{B\arcmin F}-\stackrel{\frown}{BF})|. \label{a12}
\end{eqnarray}
Assuming AC$\arcmin$ = $d\arcmin$, which similar to Eqs.\ref{a10} \& \ref{a11}, we can get
\begin{eqnarray}
 \rm \stackrel{\frown}{A\arcmin D} &=& \rm R_\sun * \arcsin(\frac{\it d \arcmin}{\rm R_\sun}) \label{a13}\\
 \rm \stackrel{\frown}{B\arcmin F} &=& \rm (R_\sun + \Delta \it h)* \beta \arcmin \label{a14},
\end{eqnarray}
where
\begin{eqnarray}
  \beta \arcmin &=& \arcsin\{\frac{\rm(R_\sun + 1 AU)*\cos(\theta \arcmin) - \rm \sqrt{(R_\sun +\Delta \it h \rm)^2-[(R_\sun + 1 AU)*\sin(\theta \arcmin)]^2}}{(\rm R_\sun + \Delta \it h)}*\sin(\theta \arcmin)\}\nonumber \\
  \theta \arcmin &=& \arctan(\frac{d\arcmin}{\rm R_\sun-\sqrt{R_\sun^2 - \it d\arcmin\rm^2} + 1 AU}).\nonumber
\end{eqnarray}
On substituting Eqs.\ref{a10}, \ref{a11}, \ref{a13}, \ref{a14} in Eq.\ref{a12}, we obtain
\begin{eqnarray}
\Delta l\arcmin&=&| \rm R_\sun*[\arcsin(\frac{d\arcmin}{R_\sun})-\arcsin(\frac{d}{R_\sun})]-(R_\sun + \Delta \it h)*(\beta \arcmin- \beta)|. \label{a15}
\end{eqnarray}

The error in velocity measurement would be
\begin{eqnarray}
  \Delta v\arcmin &=& \frac{\Delta l\arcmin}{t} \nonumber\\
           &=&\frac{\Delta l\arcmin}{\rm \stackrel{\frown}{AA\arcmin}}*\frac{\rm \stackrel{\frown}{AA\arcmin}}{t} \nonumber\\
           &=&\frac{\Delta l\arcmin}{R_\sun*[\arcsin(\frac{d\arcmin}{\rm R_\sun})-\arcsin(\frac{d}{\rm R_\sun})]}*v_{obs}, \label{a16}
\end{eqnarray}
where $v_{obs}$ is the velocity of observation. Here, R$_\sun$ = 6.963$\times 10^5$ km, 1 AU = 1.4959787$\times 10^8$ km. On the other hand, we assume that $d\arcmin$ = 0.8 R$_\sun$, $d$ = 0, $\Delta h$ = 100 Mm (the typical EUV wave height range), which means that the EUV wave propagate along the ``corresponding longitude" from the disk center to the place at 0.8 R$_\sun$. According to the Eqs.\ref{a15} \& \ref{a16}, we can get $\Delta l\arcmin \approx$29.5 Mm and $\Delta v\arcmin\approx$4.56\% $v_{obs}$. If $d\arcmin$ = 0.8 R$_\sun$, $d$ = 0.6 R$_\sun$, we can get $\Delta l\arcmin \approx$20.8 Mm and $\Delta v\arcmin\approx$10.56\% $v_{obs}$. This means that the error is significant when the wave travels along the ``corresponding longitude", especially near the solar limb.

Secondly, if the EUV wave initially propagates from B to B$\arcsec$ along the ``corresponding latitude", taking time of t, which projects onto the solar surface from A to A$\arcsec$ (see Fig.\ref{fig12} (b)), then the different distance ($\Delta l\arcsec$) could be expressed as
\begin{eqnarray}
  \Delta l\arcsec &=& |\rm \stackrel{\frown}{AA\arcsec} -\stackrel{\frown}{BB\arcsec}| \nonumber \\
           &=& |\rm \alpha*(AC - BC\arcsec)| \nonumber \\
           &=& |\rm \alpha*[AC-OB*\sin(\angle BOE)]|,\label{a17}
\end{eqnarray}
\textbf{where $\alpha =\rm \angle ACA\arcsec \equiv \angle BC\arcmin B\arcsec$}.
On substituting Eqs.\ref{a3}, \ref{a7} \& \ref{a9} in Eq.\ref{a17}, we obtain
\begin{eqnarray}
\Delta l\arcsec &=& |\alpha*[d-(\rm R_\sun+\Delta \it h)*\sin(\beta)]|. \label{a18}
\end{eqnarray}

The error in velocity measurement would be
\begin{eqnarray}
  \Delta v\arcsec &=& \frac{\Delta l\arcsec}{t} \nonumber\\
           &=&\frac{\Delta l\arcsec}{\rm \stackrel{\frown}{AA\arcsec}}*\frac{\rm \stackrel{\frown}{AA\arcsec}}{t} \nonumber\\
           &=&\frac{\Delta l\arcsec}{\alpha*d}*v_{obs}. \label{a19}
\end{eqnarray}
We also assume some parameters that $\alpha= \frac{\pi}{6}$, $d$ = 0.8 R$_\sun$, $\Delta h$ = 100 Mm. According to the Eqs.\ref{a18} \& \ref{a19}, we can get $\Delta l\arcsec\approx$ 0.3 Mm and $\Delta v\arcsec\approx$  0.10\% $v_{obs}$. If $\alpha= \frac{\pi}{2}$, $d$ = 0.8 R$_\sun$, $\Delta h$ = 100 Mm, we can get $\Delta l\arcsec\approx$ 0.89 Mm and $\Delta v\arcsec\approx$  0.10\% $v_{obs}$. Therefore, when the wave travels along the ``corresponding latitude", the error in the measurement is tiny.

Thirdly, for more reality, if the EUV wave propagates through both ``corresponding longitude" and ``corresponding latitude" from B to B$^\ast$, which projects onto the solar surface from A to A$^\ast$ (see Fig.\ref{fig12} (c)), then the different distance ($\Delta l$) could be expressed as
\begin{eqnarray}
 \Delta l = |\rm \stackrel{\frown}{AA^\ast} - \rm \stackrel{\frown}{BB^\ast}|.\label{a20}
\end{eqnarray}
We set that AC$^\ast$ = $d\arcmin$. A$^\ast$ and A$\arcmin$ are at the same ``corresponding latitude" cycle while B$^\ast$ and B$\arcmin$ are the same ``corresponding latitude" cycle. According to the Haversine Formula, we can get

\begin{eqnarray}
\rm \stackrel{\frown}{AA^\ast} &\approx& \rm R_\sun*\arccos\{1-2[haversin(\varphi_2 - \varphi_1)+cos(\varphi 1)cos(\varphi 2)haversin(\alpha)]\} \nonumber\\
 &\approx& \rm R_\sun * \arccos\{\cos[\arcsin(\frac{\it d\arcmin}{R_\sun})-\arcsin(\frac{\it d}{R_\sun})]-\frac{\it d*d\arcmin}{R^2_\sun}[1-\cos(\alpha)]\} \nonumber \\
\rm \stackrel{\frown}{BB^\ast} &\approx& \rm (R_\sun+\it \Delta h)*\arccos\{\rm 1-2[haversin(\Phi_2 - \Phi_1)+cos(\Phi_1)cos(\Phi_2)haversin( \alpha)]\} \nonumber \\
&\approx& \rm (R_\sun+\it \Delta h)*\arccos\{\cos(\beta\arcmin - \beta)-\sin(\beta\arcmin)\sin(\beta)[\rm 1-\cos(\alpha)]\}
\end{eqnarray}
where
\begin{eqnarray}
  \varphi_1 &=& \frac{\pi}{2}-\arcsin (\frac{d}{\rm R_\sun}),\nonumber \\
  \varphi_2 &=& \frac{\pi}{2}-\arcsin (\frac{d\arcmin}{\rm R_\sun}),\nonumber \\
     \Phi_1 &=& \frac{\pi}{2}-\rm \angle BOE, \nonumber \\
            &=& \frac{\pi}{2}-\beta, \nonumber \\
     \Phi_2 &=& \frac{\pi}{2}-\rm \angle B\arcmin OE ,\nonumber \\
            &=& \frac{\pi}{2}-\beta \arcmin,\nonumber \\
\rm haversin(\lambda) &=&\sin^2(\frac{\lambda}{2}),\nonumber\\
                      &=&\frac{1-\cos(\lambda)}{2}.\nonumber
\end{eqnarray}
Additionally,
\begin{eqnarray}
  \beta&=&\arcsin\{\frac{\rm(R_\sun + 1 AU)*\cos(\theta) - \rm \sqrt{(R_\sun +\Delta \it h \rm)^2-[(R_\sun + 1 AU)*\sin(\theta)]^2}}{(R_\sun + \Delta h)}*\sin(\theta)\}\nonumber \\
   \beta \arcmin &=& \arcsin\{\frac{\rm(R_\sun + 1 AU)*\cos(\theta \arcmin) - \rm \sqrt{(R_\sun +\Delta \it h \rm)^2-[(R_\sun + 1 AU)*\sin(\theta \arcmin)]^2}}{(\rm R_\sun + \Delta \it h)}*\sin(\theta \arcmin)\}\nonumber \\
  \theta &=& \arctan(\frac{d}{\rm R_\sun-\sqrt{R_\sun^2 - \it d \rm^2} + 1 AU})\nonumber\\
  \theta \arcmin &=& \arctan(\frac{d\arcmin}{\rm R_\sun-\sqrt{R_\sun^2 - \it d\arcmin\rm^2} + 1 AU}).\nonumber
\end{eqnarray}
The error in velocity measurement would be
\begin{eqnarray}
  \Delta v &=& \frac{\Delta l}{t} \nonumber\\
           &=&\frac{\Delta l}{\rm \stackrel{\frown}{AA^\ast}}*\frac{\rm \stackrel{\frown}{AA^\ast}}{t} \nonumber\\
           &=&\frac{\Delta l}{\rm R_\sun * \arccos\{\cos[\arcsin(\frac{\it d\arcmin}{R_\sun})-\arcsin(\frac{\it d}{R_\sun})]-\frac{\it d*d\arcmin}{R^2_\sun}[1-\cos(\alpha)]\}}*v_{obs}. \label{a21}
\end{eqnarray}
According to above equations, we can find that the the error in the measurement is related to $\alpha$, $d$, $d\arcmin$, $\Delta h$. The main error comes from the first part ($\Delta l\arcmin$). Based on our study event, for Sector A, we assume that $\alpha= \frac{\pi}{2}*\frac{8}{9}$, $d$ = 0.8 R$_\sun$, $d\arcmin$ = 0.9 R$_\sun$, $\Delta h$ = 100 Mm. According to the Eqs.\ref{a20} \& \ref{a21}, we can get $\Delta v\approx$  2.3\% $v_{obs}$. If we assume that $\Delta h = 50$ Mm (about half of the typical EUV wave height range), we can obtain $\Delta v\approx$  1.4\% $v_{obs}$. For Sector B, we assume that $\alpha= \frac{\pi}{2}*\frac{6}{9}$, $d$ = 0.8 R$_\sun$, $d\arcmin$ = 0.4 R$_\sun$, $\Delta$= 100 Mm, we get $\Delta v \approx$ 4.0\% $v_{obs}$. For Sector D, we assume that $\alpha= \frac{\pi}{2}*\frac{8}{9}$, $d$ = 0.8 R$_\sun$, $d\arcmin$ = 0.2 R$_\sun$, $\Delta$= 100 Mm, we get $\Delta v\approx$ 4.4\% $v_{obs}$. It should take notice that these errors are the overestimation errors in the measurement.



\begin{thebibliography}{}
\bibitem[Altschuler, \& Newkirk(1969)]{alt69} Altschuler, M.~D., \& Newkirk, G.\ 1969, \solphys, 9, 131
\bibitem[Al-Omari et al.(2010)]{alo10} Al-Omari, M., Qahwaji, R., Colak, T., et al.\ 2010, Solar Physics, 262, 511
\bibitem[Asai et al.(2012)]{asa12} Asai, A., Ishii, T.~T., Isobe, H., et al.\ 2012, \apjl, 745, L18
\bibitem[Athay \& Moreton(1961)]{ath61} Athay, R.~G., \& Moreton, G.~E.\ 1961, \apj, 133, 935
\bibitem[Balasubramaniam et al.(2010)]{bal10} Balasubramaniam, K.~S., Cliver, E.~W., Pevtsov, A., et al.\ 2010, \apj, 723, 587
\bibitem[Ballai et al.(2005)]{bal05} Ballai, I., Erd{\'e}lyi, R., \& Pint{\'e}r, B.\ 2005, \apjl, 633, L145
\bibitem[Biesecker et al.(2002)]{bie02} Biesecker, D.~A., Myers, D.~C., Thompson, B.~J., et al.\ 2002, \apj, 569, 1009
\bibitem[Brueckner et al.(1995)]{bru95} Brueckner, G.~E., Howard, R.~A., Koomen, M.~J., et al.\ 1995, Solar Physics, 162, 357
\bibitem[Cabezas et al.(2019)]{cab19} Cabezas, D.~P., Asai, A., Ichimoto, K., et al.\ 2019, \apj, 883, 32
\bibitem[Chen \& Wu(2011)]{che11} Chen, P.~F., \& Wu, Y.\ 2011, \apjl, 732, L20
\bibitem[Chen et al.(2002)]{che02} Chen, P.~F., Wu, S.~T., Shibata, K., et al.\ 2002, \apjl, 572, L99
\bibitem[Chen(2009)]{che09} Chen, P.~F.\ 2009, \apjl, 698, L112
\bibitem[Cliver(2016)]{cli16} Cliver, E.~W.\ 2016, \apj, 832, 128
\bibitem[Delaboudini{\`e}re et al.(1995)]{del95} Delaboudini{\`e}re, J.-P., Artzner, G.~E., Brunaud, J., et al.\ 1995, Solar Physics, 162, 291
\bibitem[Delann{\'e}e \& Aulanier(1999)]{del99} Delann{\'e}e, C., \& Aulanier, G.\ 1999, \solphys, 190, 107
\bibitem[Domingo et al.(1995)]{dom95} Domingo, V., Fleck, B., \& Poland, A.~I.\ 1995, Solar Physics, 162, 1
\bibitem[Duan et al.(2019)]{dua19} Duan, Y., Shen, Y., Chen, H., et al.\ 2019, \apj, 881, 132
\bibitem[Francile et al.(2013)]{fra13} Francile, C., Costa, A., Luoni, M.~L., et al.\ 2013, \aap, 552, A3
\bibitem[Francile et al.(2016)]{fra16} Francile, C., L{\'o}pez, F.~M., Cremades, H., et al.\ 2016, \solphys, 291, 3217
\bibitem[Gallagher \& Long(2011)]{gal11} Gallagher, P.~T., \& Long, D.~M.\ 2011, \ssr, 158, 365
\bibitem[Grechnev et al.(2011)]{gre11} Grechnev, V.~V., Afanasyev, A.~N., Uralov, A.~M., et al.\ 2011, \solphys, 273, 461
\bibitem[Harra \& Sterling(2003)]{har03} Harra, L.~K., \& Sterling, A.~C.\ 2003, \apj, 587, 429
\bibitem[Harvey et al.(1996)]{har96} Harvey, J.~W., Hill, F., Hubbard, R.~P., et al.\ 1996, Science, 272, 1284
\bibitem[Harvey et al.(2011)]{har11} Harvey, J.~W., Bolding, J., Clark, R., et al.\ 2011, in AAS/Solar Physics Division Abstracts 42, 1745
\bibitem[Hong et al.(2017)]{hon17} Hong, J., Jiang, Y., Yang, J., et al.\ 2017, \apj, 835, 35
\bibitem[Hudson et al.(2003)]{hud03} Hudson, H.~S., Khan, J.~I., Lemen, J.~R., et al.\ 2003, \solphys, 212, 121
\bibitem[Kienreich et al.(2013)]{kie13} Kienreich, I.~W., Muhr, N., Veronig, A.~M., et al.\ 2013, \solphys, 286, 201
\bibitem[Klassen et al.(2000)]{kla00} Klassen, A., Aurass, H., Mann, G., et al.\ 2000, \aaps, 141, 357
\bibitem[Kozarev et al.(2011)]{koz11} Kozarev, K.~A., Korreck, K.~E., Lobzin, V.~V., et al.\ 2011, \apjl, 733, L25
\bibitem[Lemen et al.(2012)]{lem12} Lemen, J.~R., Title, A.~M., Akin, D.~J., et al.\ 2012, \solphys, 275, 17
\bibitem[Li \& Zhang(2012)]{li12} Li, T., \& Zhang, J.\ 2012, \apjl, 760, L10
\bibitem[Lin \& Forbes(2000)]{lin00} Lin, J., \& Forbes, T.~G.\ 2000, Journal of Geophysical Research, 105, 2375
\bibitem[Liu et al.(2010)]{liu10} Liu, W., Nitta, N.~V., Schrijver, C.~J., et al.\ 2010, \apjl, 723, L53
\bibitem[Liu, \& Ofman(2014)]{liu14w} Liu, W., \& Ofman, L.\ 2014, \solphys, 289, 3233
\bibitem[Liu et al.(2013)]{liu13} Liu, R., Liu, C., Xu, Y., et al.\ 2013, \apj, 773, 166
\bibitem[Liu et al.(2014)]{liu14} Liu, Z., Xu, J., Gu, B.-Z., et al.\ 2014, Research in Astronomy and Astrophysics, 14, 705-718
\bibitem[Long et al.(2019)]{lon19} Long, D.~M., Jenkins, J., \& Valori, G.\ 2019, \apj, 882, 90
\bibitem[Low(1996)]{low96} Low, B.~C.\ 1996, \solphys, 167, 217
\bibitem[Ma et al.(2011)]{ma11} Ma, S., Raymond, J.~C., Golub, L., et al.\ 2011, \apj, 738, 160
\bibitem[Mei et al.(2012a)]{mei12a} Mei, Z., Shen, C., Wu, N., et al.\ 2012, \mnras, 425, 2824
\bibitem[Mei et al.(2012b)]{mei12} Mei, Z., Udo, Z., \& Lin, J.\ 2012, Science China Physics, Mechanics, and Astronomy, 55, 1316
\bibitem[Moore et al.(2010)]{moo10} Moore, R.~L., Cirtain, J.~W., Sterling, A.~C., et al.\ 2010, \apj, 720, 757
\bibitem[Moreton(1960)]{mor60} Moreton, G.~E.\ 1960, \aj, 65, 494
\bibitem[Moreton, \& Ramsey(1960)]{more60} Moreton, G.~E., \& Ramsey, H.~E.\ 1960, \pasp, 72, 357
\bibitem[Moses et al.(1997)]{mos97} Moses, D., Clette, F., Delaboudini{\`e}re, J.-P., et al.\ 1997, \solphys, 175, 571
\bibitem[Munro et al.(1979)]{mun79} Munro, R.~H., Gosling, J.~T., Hildner, E., et al.\ 1979, Solar Physics, 61, 201
\bibitem[Ofman \& Thompson(2002)]{ofm02} Ofman, L., \& Thompson, B.~J.\ 2002, \apj, 574, 440
\bibitem[Otsuji et al.(2007)]{ots07} Otsuji, K., Shibata, K., Kitai, R., et al.\ 2007, Publications of the Astronomical Society of Japan, 59, S649
\bibitem[Parker(1963)]{par63} Parker, E.~N.\ 1963, \apjs, 8, 177
\bibitem[Pesnell et al.(2012)]{pes12} Pesnell, W.~D., Thompson, B.~J., \& Chamberlin, P.~C.\ 2012, \solphys, 275, 3
\bibitem[Priest(2014)]{pri14} Priest, E.\ 2014, Magnetohydrodynamics of the Sun
\bibitem[Raouafi et al.(2016)]{rao16} Raouafi, N.~E., Patsourakos, S., Pariat, E., et al.\ 2016, \ssr, 201, 1
\bibitem[Schatten et al.(1969)]{sch69} Schatten, K.~H., Wilcox, J.~M., \& Ness, N.~F.\ 1969, \solphys, 6, 442
\bibitem[Schmidt, \& Ofman(2010)]{sch10} Schmidt, J.~M., \& Ofman, L.\ 2010, \apj, 713, 1008
\bibitem[Schou et al.(2012)]{sch12} Schou, J., Scherrer, P.~H., Bush, R.~I., et al.\ 2012, Solar Physics, 275, 229
\bibitem[Sedov(1959)]{sed59} Sedov, L.~I.\ 1959, Similarity and Dimensional Methods in Mechanics
\bibitem[Shen, \& Liu(2012)]{she12} Shen, Y., \& Liu, Y.\ 2012, \apjl, 752, L23
\bibitem[Shen et al.(2018)]{she18} Shen, Y., Liu, Y., Liu, Y.~D., et al.\ 2018, \apj, 861, 105
\bibitem[Temmer et al.(2009)]{tem09} Temmer, M., Vr{\v{s}}nak, B., {\v{Z}}ic, T., et al.\ 2009, \apj, 702, 1343
\bibitem[Thompson et al.(1998)]{tho98} Thompson, B.~J., Plunkett, S.~P., Gurman, J.~B., et al.\ 1998, \grl, 25, 2465
\bibitem[Thompson, \& Myers(2009)]{tho09} Thompson, B.~J., \& Myers, D.~C.\ 2009, \apjs, 183, 225
\bibitem[Tripathi et al.(2009)]{tri09} Tripathi, D., Isobe, H., \& Jain, R.\ 2009, \ssr, 149, 283
\bibitem[Tripathi, \& Raouafi(2007)]{tri07} Tripathi, D., \& Raouafi, N.-E.\ 2007, \aap, 473, 951
\bibitem[Uchida(1968)]{uch68} Uchida, Y.\ 1968, \solphys, 4, 30
\bibitem[Uchida et al.(1973)]{uch73} Uchida, Y., Altschuler, M.~D., \& Newkirk, G.\ 1973, \solphys, 28, 495
\bibitem[Uralov et al.(2019)]{ura19} Uralov, A.~M., Grechnev, V.~V., \& Ivanukin, L.~A.\ 2019, \solphys, 294, 113
\bibitem[Veronig et al.(2008)]{ver08} Veronig, A.~M., Temmer, M., \& Vr{\v{s}}nak, B.\ 2008, \apjl, 681, L113
\bibitem[Vr{\v{s}}nak et al.(2002)]{vrs02} Vr{\v{s}}nak, B., Warmuth, A., Braj{\v{s}}a, R., et al.\ 2002, \aap, 394, 299
\bibitem[Vr{\v{s}}nak et al.(2006)]{vrs06} Vr{\v{s}}nak, B., Warmuth, A., Temmer, M., et al.\ 2006, \aap, 448, 739
\bibitem[Vr{\v{s}}nak, \& Cliver(2008)]{vrs08} Vr{\v{s}}nak, B., \& Cliver, E.~W.\ 2008, \solphys, 253, 215
\bibitem[Vr{\v{s}}nak et al.(2016)]{vrs16} Vr{\v{s}}nak, B., {\v{Z}}ic, T., Luli{\'c}, S., et al.\ 2016, \solphys, 291, 89
\bibitem[Wang et al.(2009)]{wan09} Wang, H., Shen, C., \& Lin, J.\ 2009, \apj, 700, 1716
\bibitem[Wang et al.(2017)]{wan17} Wang, J., Yan, X., Qu, Z., et al.\ 2017, \apj, 839, 128
\bibitem[Wang et al.(2018)]{wan18} Wang, J., Yan, X., Qu, Z., et al.\ 2018, \apj, 863, 180
\bibitem[Wang(2000)]{wan00} Wang, Y.-M.\ 2000, \apjl, 543, L89
\bibitem[Warmuth et al.(2004)]{war04} Warmuth, A., Vr{\v{s}}nak, B., Magdaleni{\'c}, J., et al.\ 2004, \aap, 418, 1117
\bibitem[Warmuth et al.(2005)]{war05} Warmuth, A., Mann, G., \& Aurass, H.\ 2005, \apjl, 626, L121
\bibitem[Warmuth, \& Mann(2011)]{war11} Warmuth, A., \& Mann, G.\ 2011, \aap, 532, A151
\bibitem[Warmuth(2015)]{war15} Warmuth, A.\ 2015, Living Reviews in Solar Physics, 12, 3
\bibitem[Wills-Davey(2006)]{wil06} Wills-Davey, M.~J.\ 2006, \apj, 645, 757
\bibitem[Wu et al.(2001)]{wu01} Wu, S.~T., Zheng, H., Wang, S., et al.\ 2001, Journal of Geophysical Research, 106, 25089
\bibitem[Xiang et al.(2016)]{xia16} Xiang, Y.-. yuan ., Liu, Z., \& Jin, Z.-. yu .\ 2016, \na, 49, 8
\bibitem[Xie et al.(2019)]{xie19} Xie, X., Mei, Z., Huang, M., et al.\ 2019, \mnras, 490, 2918
\bibitem[Xu et al.(2019)]{xu19} Xu, Z., Yang, J., Ji, K., et al.\ 2019, \apj, 874, 134
\bibitem[Xue et al.(2013)]{xue13} Xue, Z.~K., Qu, Z.~Q., Yan, X.~L., et al.\ 2013, \aap, 556, A152
\bibitem[Xue et al.(2016)]{xue16} Xue, Z., Yan, X., Cheng, X., et al.\ 2016, Nature Communications, 7, 11837
\bibitem[Yan et al.(2017)]{yan17} Yan, X.~L., Jiang, C.~W., Xue, Z.~K., et al.\ 2017, \apj, 845, 18
\bibitem[Yan et al.(2018)]{yan18} Yan, X.~L., Yang, L.~H., Xue, Z.~K., et al.\ 2018, \apjl, 853, L18
\bibitem[Yan et al.(2019)]{yan19} Yan, X., Liu, Z., Zhang, J., et al.\ 2019, arXiv e-prints, arXiv:1910.09127
\bibitem[Zhang et al.(2001)]{zha01} Zhang, J., Dere, K.~P., Howard, R.~A., et al.\ 2001, \apj, 559, 452
\bibitem[Zhang et al.(2011)]{zha11} Zhang, Y., Kitai, R., Narukage, N., et al.\ 2011, \pasj, 63, 685
\bibitem[Zhang et al.(2018)]{zha18} Zhang, J., Tian, H., Solanki, S.~K., et al.\ 2018, \apj, 865, 29
\bibitem[Zheng et al.(2016)]{zhe16} Zheng, R., Chen, Y., Du, G., et al.\ 2016, \apjl, 819, L18
\bibitem[Zheng et al.(2017)]{zhe17} Zheng, R., Zhang, Q., Chen, Y., et al.\ 2017, \apj, 836, 160

\end{thebibliography}
\end{document}